\documentclass[12pt, draftclsnofoot, onecolumn]{IEEEtran}

\usepackage[english]{babel}
\usepackage{blindtext}
\usepackage{enumitem}

\usepackage{amsfonts,amssymb,bm}
\usepackage{graphicx}
\usepackage{algorithm, algorithmicx}
\usepackage[noend]{algpseudocode}
\usepackage{amsmath}
\usepackage{amsthm}
\usepackage{cite}  % [1],[2],[3] --> [1]-[3]
\makeatletter
\renewcommand{\fnum@figure}{Fig. \thefigure}
\makeatother

\usepackage{xcolor}

\newcommand{\orange}[1]{\textcolor{black}{#1}}
\newcommand{\neworange}[1]{\textcolor{black}{#1}}

 \usepackage{mathtools}

\makeatletter
\renewcommand{\paragraph}{\@startsection{paragraph}{4}{5ex}%
   {-3.25ex plus -1ex minus -0.2ex}%
   {1.5ex plus 0.2ex}%
   {\normalfont\normalsize\bfseries}}
\makeatother

\stepcounter{secnumdepth}
\stepcounter{tocdepth}

\renewcommand{\figurename}{Fig.}

\ifCLASSINFOpdf
\else
\fi
\usepackage{amsmath}
\ifCLASSOPTIONcompsoc
 \usepackage[caption=false,font=normalsize,labelfont=sf,textfont=sf]{subfig}
\else
 \usepackage[caption=false,font=footnotesize]{subfig}
\fi
\begin{document}
\begin{sloppypar}
% \captionsetup[figure]{labelfont={bf},labelformat={default},labelsep=period,name={Fig.}}
% \captionsetup[figure]{name={Fig.}}

%
% paper title
% Titles are generally capitalized except for words such as a, an, and, as,
% at, but, by, for, in, nor, of, on, or, the, to and up, which are usually
% not capitalized unless they are the first or last word of the title.
% Linebreaks \\ can be used within to get better formatting as desired.
% Do not put math or special symbols in the title.
% \title{ENGNN: A New Deep Learning Architecture for Wireless Beamforming Problems}

\title{ENGNN: A General Edge-Update Empowered GNN Architecture for Radio Resource Management in Wireless Networks}

% Beamforming Design

% Graph
% Edge-Update Mechanism
% ENGNN

%
%
% author names and IEEE memberships
% note positions of commas and nonbreaking spaces ( ~ ) LaTeX will not break
% a structure at a ~ so this keeps an author's name from being broken across
% two lines.
% use \thanks{} to gain access to the first footnote area
% a separate \thanks must be used for each paragraph as LaTeX2e's \thanks
% was not built to handle multiple paragraphs
%

% \author{Yunqi Wang, %~\IEEEmembership{Member,~IEEE,}
%         Yang Li, %~\IEEEmembership{Fellow,~OSA,}
%         Qingjiang Shi, %~
%         and Yik-Chung~Wu%,~\IEEEmembership{Life~Fellow,~IEEE}% <-this % stops a space
% % \thanks{M. Shell was with the Department
% % of Electrical and Computer Engineering, Georgia Institute of Technology, Atlanta,
% % GA, 30332 USA e-mail: (see http://www.michaelshell.org/contact.html).}% <-this % stops a space
% % \thanks{J. Doe and J. Doe are with Anonymous University.}% <-this % stops a space
% % \thanks{Manuscript received April 19, 2005; revised August 26, 2015.}
% }

\author{Yunqi Wang,
        Yang Li,
        Qingjiang Shi,
        and Yik-Chung Wu

\thanks{Yunqi Wang is with the Department of Electrical and Electronic Engineering, The University of Hong Kong, Hong Kong, 
and also with Shenzhen Research Institute of Big Data, Shenzhen 518172, China (email: yunqi9@connect.hku.hk).}
\thanks{Yang Li is with Shenzhen Research Institute of Big Data, Shenzhen 518172, China (e-mail: liyang@sribd.cn).}
\thanks{Qingjiang Shi is with the School of Software Engineering, Tongji University, Shanghai 200092, China, 
and also with Shenzhen Research Institute of Big Data, Shenzhen 518172, China (email: shiqj@tongji.edu.cn). }
\thanks{Yik-Chung Wu is with the Department of Electrical and Electronic Engineering, The University of Hong Kong, Hong Kong 
(email: ycwu@eee.hku.hk).} 
}

\maketitle

% As a general rule, do not put math, special symbols or citations
% in the abstract or keywords.
\begin{abstract}
In order to achieve high data rate and ubiquitous connectivity in future wireless networks, a key task is to 
efficiently manage the radio resource by judicious beamforming and power allocation. 
Unfortunately, the iterative nature of the commonly applied optimization-based algorithms 
cannot meet the low latency requirements due to the high computational complexity.
For real-time implementations, deep learning-based approaches, especially the graph neural networks (GNNs),
have been demonstrated with good scalability and generalization performance due to the permutation equivariance (PE)
property. However, the current architectures are only equipped with the node-update mechanism, 
which prohibits the applications to a more general setup, where the unknown variables 
are also defined on the graph edges. To fill this gap, we propose an edge-update mechanism, 
which enables GNNs to handle both node and edge variables and prove its PE property with respect to both transmitters and receivers. 
Simulation results on typical radio resource management problems demonstrate that the proposed method achieves higher sum rate but with much shorter computation time than state-of-the-art methods and generalizes well on different numbers of base stations and users, different noise variances, interference levels, and transmit power budgets.
\end{abstract}

% Note that keywords are not normally used for peerreview papers.
\begin{IEEEkeywords}
Beamforming design, power allocation, heterogeneous graph neural network (GNN), edge-update
mechanism. 
\end{IEEEkeywords}

% For peer review papers, you can put extra information on the cover
% page as needed:
% \ifCLASSOPTIONpeerreview
% \begin{center} \bfseries EDICS Category: 3-BBND \end{center}
% \fi
%
% For peerreview papers, this IEEEtran command inserts a page break and
% creates the second title. It will be ignored for other modes.
\IEEEpeerreviewmaketitle

% \section{Introduction}
% The very first letter is a 2 line initial drop letter followed
% by the rest of the first word in caps.
% 
% form to use if the first word consists of a single letter:
% \IEEEPARstart{A}{demo} file is ....
% 
% form to use if you need the single drop letter followed by
% normal text (unknown if ever used by the IEEE):
% \IEEEPARstart{A}{}demo file is ....
% 
% Some journals put the first two words in caps:
% \IEEEPARstart{T}{his demo} file is ....
% 
% Here we have the typical use of a "T" for an initial drop letter
% and "HIS" in caps to complete the first word.

% \IEEEPARstart{W}{ireless} systems often have difficulty implementing beamforming policies in real time given time-varying channels. Moreover, these problems are often non-convex and computationally expensive, making it more challenging to obtain efficient solutions.

% ######################################################################################### %
%                                    Section I   Introduction                               %
% ######################################################################################### %
\section{Introduction}

Efficient radio resource management
plays a vital role in achieving high data rate and ubiquitous connectivity of future wireless networks.
In particular, beamforming design and power allocation 
have been recognized as crucial components to improve the spectrum/energy efficiency in ultra-dense networks\cite{zhang2004cochannel}, cloud radio access networks \cite{shi2015large, li2018first}, 
and cell-free massive multiple-input multiple-output systems \cite{he2021cell}.

Mathematically, many of the radio resource management problems belong to the challenging non-convex optimization problems, which are
conventionally solved by numerical algorithms 
with a lot of iterations \cite{li2019energy, mohammadi2022fast}. 
However, 
due to the fast variation of the wireless
environment, 
the iterative nature of the commonly applied optimization-based numerical algorithms cannot satisfy the low-latency requirement in beyond 5G paradigm. 
For instance, to maximize the sum rate of a multi-cell 
wireless system under the maximum
transmit power constraint of each base station (BS),
the conventional first-order algorithm, e.g., 
the gradient projection (GP) method~\cite{bertsekas1997nonlinear}, requires a lot of iterations to converge to a stationary point.
To improve the convergence rate,
while more advanced numerical algorithms such as 
the weighted minimum mean-square error (WMMSE) \cite{shi2011iteratively} algorithm can be applied,
the matrix inverse in each iteration still
makes it computationally expensive and hence difficult 
for the real-time implementation.

To facilitate the real-time implementation,
deep learning based methods have become popular for radio resource management \cite{ 
sun2018learning,  
liang2019towards,
lee2018deep, 
xia2019deep, 
cui2019spatial, 
shen2019lorm, 
zhu2020learning,
ma2021neural}.
Specifically, deep learning based methods utilize neural networks to  
learn a mapping function from many problem features to the corresponding solutions. 
Once the neural network is well trained, 
it can infer the solution of any new setting using simple feed-forward computations, and thus is extremely fast.

Inspired by the successful applications in
computer vision, the multi-layer perceptrons (MLPs) and convolutional neural networks (CNNs)
have been applied as typical architectures for representing the mapping functions in
radio resource management.  
For example, MLPs were used to learn the mapping function
from the wireless channel to the optimal resource management policy \cite{sun2018learning}. 
Moreover, an MLP-based architecture was adopted to 
learn the optimal power control for the multi-user interference channels \cite{liang2019towards}. 
Similarly, CNNs have also been applied for power control \cite{lee2018deep} and 
beamforming design \cite{xia2019deep} in the multiple-input single-output downlink systems. 
However, since MLPs and CNNs cannot fully exploit the topology in the wireless networks, 
they usually require a large number of training samples 
while still result in limited 
performance.
For instance, it is shown in \cite{ma2021neural} that 
a CNN trained on a two-user wireless networks can only achieve the
near-optimal performance for two-user wireless networks during the 
testing phase, 
but its performance degrades by $18\%$ in ten-user networks compared to the conventional 
optimization-based numerical algorithms.

Recently, attempts to use graph neural networks (GNNs) are on the rise because of their ability to exploit the topology of wireless networks.
By modeling a wireless network as a graph, the known
system parameters can be modeled as the graph features, which are treated as the input of a GNN,
while the unknown variables to be optimized can also be defined on the graph
and are served as the output of a GNN.
The advantage of graph modeling lies in its permutation equivariance (PE) property, 
where the graph features/variables can be regarded as a set of elements whose index order does not matter. 
Consequently, a large number of unnecessary permuted training samples can be discarded~\cite{shen2020graph, 
eisen2020optimal, shen2022neural}. 
Moreover, since the trainable parameters of GNNs are independent of the graph size, 
the well-trained GNNs can generalize well to different problem dimensions~\cite{shen2021ai, li2022heterogeneous, li2021learning, shen2022graph, zhang2022learning}.

Among the existing GNN-based works, homogeneous GNNs,
which share the trainable parameters among different graph nodes,
have shown their good scalability and generalization
performance for the radio resource management problems when there is
only one type of graph nodes \cite{
shen2020graph, 
eisen2020optimal, 
shen2022neural, 
shen2021ai, 
zhang2022learning, 
lee2020graph, 
chowdhury2021unfolding,
nikoloska2022modular, 
he2022gblinks}.
For example, in \cite{shen2020graph},
a homogeneous GNN named message passing graph
neural network (MPGNN) was proposed for beamforming design in the 
multi-transceiver interference channels,
where each transceiver pair is modeled as an individual graph node.
By modeling different transceiver pairs as the same type of nodes and sharing
their trainable parameters, the test performance in terms of sum rate is near optimal even when the number of 
transceiver pair is twice larger than that in the training samples.
Similarly, homogeneous GNNs have also been demonstrated to generalize well on
different numbers of users in multicast beamforming design \cite{zhang2022learning}, 
% [Zhe Zhang, Meixia Tao, and Ya-Feng Liu],
link scheduling \cite{lee2020graph}, power control \cite{chowdhury2021unfolding, nikoloska2022modular}, and joint beam selection and link activation \cite{he2022gblinks}.

While homogeneous GNNs have shown their great success 
when there exists only one type of graph nodes in the radio resource management problems,
it should be noticed that the more common scenarios usually consist of different types of graph nodes. 
For instance, in a general wireless network, the transmitters and receivers have different 
physical characteristics, and hence should be more naturally modeled as two different types of graph nodes,
i.e., TX-nodes and RX-nodes. 
By sharing the trainable parameters within only the same type of nodes, heterogeneous GNNs \cite{zhang2019heterogeneous} have shown their superiority for the more complex radio resource management problems compared with the homogeneous GNNs 
\cite{guo2021learning,zhang2021scalable, kim2022bipartite,jiang2021learning, zhang2022cooperative}.
In particular, the pioneer work \cite{guo2021learning} designed a heterogeneous GNN 
called permutation equivariant heterogeneous GNN (PGNN) 
for the power allocation in multi-cell downlink systems, and theoretically established
the PE property with respect to different user equipments (UEs) within each cell and 
also with respect to different cells.
Moreover, heterogeneous GNNs have also been proposed for 
the beamforming design in heterogeneous device-to-device networks
\cite{zhang2021scalable} and multi-user downlink systems \cite{kim2022bipartite}. 
In \cite{jiang2021learning}, a heterogeneous GNN was designed for jointly learning the 
beamforming vectors and reflecting phases for 
an intelligent reflecting surfaces (IRS) assisted multi-user downlink system,
where the users and the IRS are modeled as heterogeneous graph nodes.
Similarly, the trajectory of unmanned aerial vehicles (UAVs)
was cooperatively designed by modeling the UAVs and the ground terminals as
heterogeneous graph nodes in \cite{zhang2022cooperative}.

Despite the successes of the homogeneous or heterogeneous GNNs in the above existing works,
they are only equipped with the node-update mechanism, which restricts
the output of the neural networks, i.e., the unknown variables to be optimized only appear on the graph nodes.
Notable examples are MPGNN~\cite{shen2020graph} and PGNN~\cite{guo2021learning}, both of which do not 
consider the variables on the graph edges.
In particular, MPGNN is proposed for the beamforming design 
for the multi-transceiver interference channels, where each transmitter only serves a single receiver. 
Therefore, each transceiver pair 
can be modeled as a single node, and the channel state information of each direct communication link
serves as the corresponding node features.
Furthermore, the interference links among different transceiver pairs 
are modeled as graph edges, whose channel state information
is treated as the corresponding edge features.
Using this graph model, the beamforming variable of each transceiver pair can be defined only on the corresponding node.
On the other hand, PGNN is proposed for the power allocation in multi-cell systems, 
where each BS serves multiple UEs within the same cell. In this scenario,
each BS adopts a pre-designed beamformer, so that a dedicated equivalent single-antenna channel is created for each UE.
Consequently, with each equivalent transmit antenna treated as an individual node, 
the power allocation variables can also be defined on the graph nodes.

While all the above pioneering works 
exemplify 
the benefits of GNNs in radio resource management, 
currently applied architectures prohibit the extension to a more general setting,
where the unknown variables are also defined on the graph edges.
A typical application scenario is the cooperative beamforming design, 
where each transmitter serves multiple receivers, while each receiver is also served by multiple transmitters. 
These complicated transceiver interactions cannot be easily modeled 
by the current GNN architectures that are only equipped with the node-update mechanism.
In fact, for the more general wireless environment,
an individual beamforming or power variable belongs to a transceiver pair, 
which is represented by two different nodes. Thus, beamformers or power variables 
should be more naturally defined 
on the graph edges. Unfortunately, without a judiciously designed edge-update mechanism,
the current widely adopted GNN architectures cannot handle such a general setting.

To fill this gap, we propose a novel edge-update mechanism, which enables the GNN
architecture to deal with both the edge and node variables for the radio resource management problems.
The contributions of this paper are summarized as follows.
\begin{enumerate}
\item We propose a general problem formulation using the heterogeneous graph for the radio resource management problems,
where the unknown variables to be optimized can be defined on the graph edges.
To learn the edge variables, we design a novel edge-update mechanism
and prove its PE property with respect to both the transmitters and receivers.
Compared with the existing node-update mechanism that gathers
the information from the neighboring nodes, the update of an edge variable
is more challenging, since it is more complicated to define the neighbors of an edge, 
let alone how to aggregate their representations.
Based on the observation that the neighboring edges can be
divided into two categories according to their connected nodes,
we propose an edge-update mechanism that extracts the
information from the two types of neighboring edges in a different manner.  
\item Based on the edge-update mechanism, we propose an edge-update empowered
neural network architecture termed as edge-node GNN (ENGNN), which can represent the mapping function from the graph features
to the edge/node variables for the radio resource management problems.
We prove that the proposed ENGNN is permutation equivariant with respect to both transmitters and receivers.
Moreover, since the trainable parameters of the proposed ENGNN are independent of the 
graph size, it can generalize to different numbers of transmitters and receivers.
Last but not the least, the proposed ENGNN can be applied in a wide range of radio resource management problems, where 
the variables occur between any pair of the TX-nodes and RX-nodes.
\item Simulation results demonstrate the superiority of the proposed ENGNN for
typical radio resource management problems, including the beamforming design in the interference channels, 
the power allocation in the interference broadcast channels, 
and the cooperative beamforming design, respectively. 
It is shown that the proposed ENGNN achieves higher sum rate with much shorter computation time than
state-of-the-art methods and generalizes well on different numbers of BSs and UEs, different noise variances,
interference levels, and transmit power budgets.
\end{enumerate}

Notations: 
In this paper, we use bold lowercase letters, bold uppercase letters, and bold italicized uppercase letters to represent vectors, matrices, and tensors, respectively. The sets are represented by stylized uppercase letters. 
The notations $(\cdot)^T$ and $(\cdot)^H$ refer to transpose and Hermitian transpose, respectively.
Moreover, $|\cdot|^2$ denotes the $l_2$-norm operation, and $|\cdot|$ computes the magnitude of a complex number or the cardinality of a set.

The rest of the paper is organized as follows. 
In Section~\ref{sec: wireless network as HetGraph}, we propose a general problem formulation on the heterogeneous graph for the radio resource management problems. 
In Section~\ref{sec:ENGNN}, we design a novel neural network
architecture named ENGNN with both the edge-update and node-update mechanisms. 
Then, in Section~\ref{sec: simulation-results}, numerical results are presented to demonstrate the superiority of the proposed ENGNN on three typical scenarios. 
Finally, the conclusion is drawn in Section~\ref{sec: concluion}.

% ######################################################################################### %
%                Section II     wireless network as a graph                  %
% ######################################################################################### %
\section{Problem Formulation on Heterogeneous Graph}
\label{sec: wireless network as HetGraph}

\subsection{General Graph Modeling}

Consider a wireless network with $M$ transmitters and $K$ receivers, which can be modeled by a heterogeneous graph. 
Specifically, the transmitters and receivers can be viewed as two types of nodes, i.e., TX-nodes and RX-nodes, respectively.
Moreover, an edge is drawn between a TX-node and an RX-node if there exists a direct communication or interference link between them. 
Such a heterogeneous graph can be expressed as $\mathcal{G} = \left\{\mathcal{M}, \mathcal{K}, \mathcal{E} \right\}$, where $\mathcal{M}\triangleq\{1,\ldots,M\}$ is the set of TX-nodes, 
$\mathcal{K}\triangleq\{1,\ldots,K\}$ is the set of RX-nodes, 
and $\mathcal{E}\subseteq\left\{(m,k)\right\}_{m\in\mathcal{M},k\in\mathcal{K}}$ is the set of edges, 
respectively. 

The TX-nodes, RX-nodes, and edges may contain features and/or variables. 
Specifically, features are known system parameters. 
For example, features on the nodes can be position coordinates, maximum transmit power budgets, and/or noise variances, 
while features on the edges can be channel state information and/or indicators of direct communication or interference links. 
On the other hand, variables are unknown beamformers and/or allocated powers to be designed.

\subsection{Problem Formulation}

Denote the feature vectors on the $m$-th TX-node and $k$-th RX-node as 
$\mathbf{f}_{\text{TX},m} \in \mathbb{C}^{d_{\text{TX}} }$ and 
$\mathbf{f}_{\text{RX},k} \in \mathbb{C}^{d_{\text{RX}} }$, where $d_{\text{TX}}$ and $d_{\text{RX}}$ denote the corresponding feature dimensions, respectively. 
Consequently, the feature matrices of TX-nodes
and RX-nodes can be expressed as 
$\mathbf{F}_{\text{TX}} = \left[\mathbf{f}_{\text{TX},1},\cdots,\mathbf{f}_{\text{TX},M}\right]^T\in\mathbb{C}^{M\times d_{\text{TX}} }$ and 
$\mathbf{F}_{\text{RX}} = \left[\mathbf{f}_{\text{RX},1},\cdots,\mathbf{f}_{\text{RX},K}\right]^T\in \mathbb{C}^{K\times d_{\text{RX}} }$. 
Similarly, we can express the edge features as a tensor $\boldsymbol{E}\in\mathbb{C}^{M\times K \times d_{\text{E}} }$, where
$d_{\text{E}}$ is the edge feature dimension.
In particular, the $(m,k,:)$-th fiber, 
$\boldsymbol{E}_{(m,k,:)}$, 
takes values if $(m,k) \in \mathcal{E}$, and is an all-zero vector otherwise. 
\label{sec: problem formulation on HetGraph}
\orange{On the other hand, 
the variables on TX-nodes and RX-nodes can be expressed as 
$\mathbf{S}_{\text{TX}} = \left[\mathbf{s}_{\text{TX},1},\cdots,\mathbf{s}_{\text{TX},M}\right]^T$ and 
$\mathbf{S}_{\text{RX}} = \left[\mathbf{s}_{\text{RX},1},\cdots,\mathbf{s}_{\text{RX},K}\right]^T$, where 
$\mathbf{s}_{\text{TX},m}\in\mathbb{C}^{d'_{\text{TX}} }$ and 
$\mathbf{s}_{\text{RX},k}\in\mathbb{C}^{d'_{\text{RX}} }$ denote the variables on the $m$-th TX-node and $k$-th RX-node, 
$d'_{\text{TX}}$ and $d'_{\text{RX}}$ denote the corresponding variable dimensions, respectively.}  
Similarly, edge variables can be expressed as $\boldsymbol{\it{\Xi}} \in \mathbb{C}^{M\times K \times d'_\text{E}}$. 

\neworange{Based on the above notations, 
a beamforming design or power allocation problem can be formulated as 
\begin{subequations}\label{equation-opt}
\begin{align}
    &\mathop{\max}\limits_{ \phi (\cdot,\cdot,\cdot)  }\ \  
    f\left( \mathbf{S}_{\text{TX}}, \mathbf{S}_{\text{RX}}, \boldsymbol{\it{\Xi}} ; \mathbf{F}_{\text{TX}}, \mathbf{F}_{\text{RX}},  \boldsymbol{E}\right), \label{equation-opt_cost}\\
    &~\text{s.t.}\ \  ~~\left(\mathbf{S}_{\text{TX}}, \mathbf{S}_{\text{RX}}, \boldsymbol{\it{\Xi}} \right) = \phi\left(\mathbf{F}_{\text{TX}}, \mathbf{F}_{\text{RX}},  \boldsymbol{E}\right), \label{equation-opt_cons2}
\end{align}
\end{subequations}
where \eqref{equation-opt_cost} is the utility function, and $\phi(\cdot,\cdot,\cdot)$ denotes the mapping function
from the features $(\mathbf{F}_{\text{TX}}, \mathbf{F}_{\text{RX}}, \boldsymbol{E})$ to 
the variables $\left(\mathbf{S}_{\text{TX}}, \mathbf{S}_{\text{RX}}, \boldsymbol{\it{\Xi}} \right)$.} 
Next, we show three typical examples under the general problem formulation \eqref{equation-opt}.

\label{sec:examples of the problems}
\emph{\textbf{Example 1: }Beamforming Design for Interference Channels.}
Consider a wireless network with $K$ BS-UE pairs, 
where the $k$-th UE is served by the $m_1(k)$-th BS, 
and $m_1(\cdot)$ is any one-to-one mapping from $\mathcal{K}$ to $\mathcal{M}$. 
Each BS is equipped with $N$ antennas, and each UE is equipped with a single antenna. 
The beamforming vector 
of the $k$-th UE 
is denoted as  $\mathbf{v}_{k}\in\mathbb{C}^{N}$, 
while the channel between 
the $m_1(k')$-th BS and the $k$-th UE 
is denoted as 
$\mathbf{h}_{m_1(k'),k} \in \mathbb{C}^{N}$. 
Then the received signal at the $k$-th UE is given by
\begin{equation}
    y_k = \mathbf{h}^H_{m_1(k),k}\mathbf{v}_{k} s_{k} + 
    \sum^K_{k'=1,k'\neq k}\mathbf{h}^H_{m_1(k'),k}\mathbf{v}_{k'} s_{k'} 
    + n_k
    ,~\forall k\in\mathcal{K},
\label{equation-exp1 received signal at RX m}
\end{equation}
where $s_k$ is the desired symbol of the $k$-th UE, 
and $n_k\sim\mathcal{C}\mathcal{N}(0,\sigma_k^2)$ is the additive complex Gaussian noise. 
Consequently, the signal-to-interference-plus-noise ratio (SINR) at the $k$-th UE can be written as 
\begin{equation}
    \text{SINR}_k = \frac{\left|\mathbf{h}^H_{m_1(k),k}\mathbf{v}_{k}\right|^2}{\sum^K_{k'=1,k'\neq k}\left|\mathbf{h}^H_{m_1(k'),k}\mathbf{v}_{k'}\right|^2+\sigma^2_k},~\forall k\in\mathcal{K}. 
\label{equation-ex1 SINR}
\end{equation}
The beamforming design problem for sum rate maximization can be formulated as: 
\begin{subequations}\label{equation-ex1 opt orig}
\begin{align}
&\mathop{\max}\limits_{ \left\{ \mathbf{v}_k \right\}_{k\in\mathcal{K}}  }~
\sum^K_{k=1} \log_2 \left(1+\text{SINR}_k\right), \label{equation-ex1 opt loss orig} \\
&~~\text{s.t.}~~~~ \left\lVert \mathbf{v}_k \right\rVert^2 \leq P_{m_1(k)},~\forall k\in\mathcal{K}, \label{equation-ex1 opt cons orig}
\end{align}
\end{subequations}
where \eqref{equation-ex1 opt cons orig} represents the maximum transmit power constraint at each BS. 

\begin{figure*}[!t]
\centering
\subfloat[Example 1: beamforming design for interference channels.]{\includegraphics[width=0.32\linewidth]{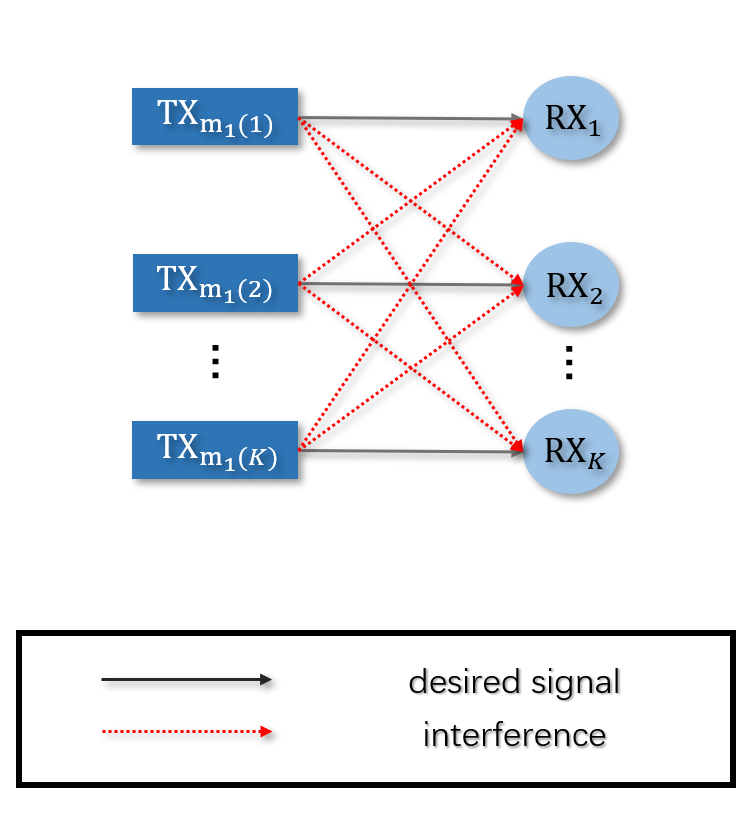}
\label{fig-ex1}}
\hfil
\subfloat[Example 2: power allocation for interference broadcast channels.]{\includegraphics[width=0.32\linewidth]{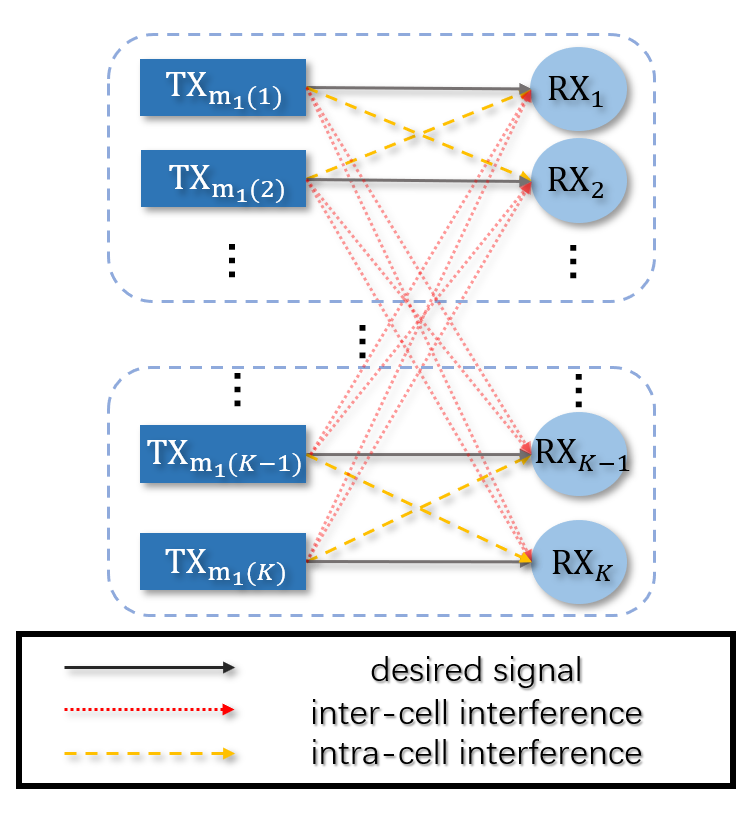}
\label{fig-ex2}}
\hfil
\subfloat[Example 3: cooperative beamforming design.]{\includegraphics[width=0.32\linewidth]{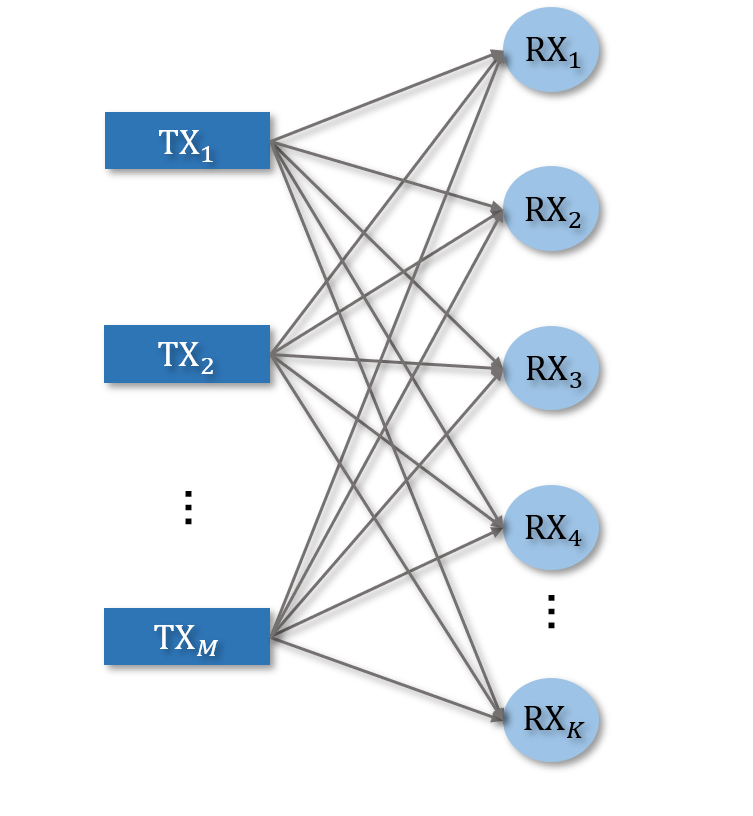}
\label{fig-ex3}}
\caption{\neworange{Graph modeling for three typical examples.}}
\label{fig-exs}
\end{figure*}

As shown in Fig.~\ref{fig-exs}\subref{fig-ex1}, 
by modeling the BSs and UEs as TX-nodes and RX-nodes respectively, we can incorporate the maximum transmit power budget $\mathbf{p} = [P_1, \cdots, P_K]^T$ and the noise standard deviation $\bm{\sigma} = [\sigma_1, \cdots, \sigma_K]^T$ as the features on the TX-nodes and RX-nodes, respectively. 
Moreover, the feature vector on the $(m,k)$-th edge contains both the channel state information $\mathbf{h}_{m,k}$ and the indicator of direct communication or interference link: 
\begin{eqnarray}\label{equation-ex1 graph feature}
\boldsymbol{H}_{(m,k,:)} = 
\begin{cases}
    [\mathbf{h}_{m,k}^T,\mathbf{0}^T]^T,~\text{if}~ m=m_1(k), \cr
    [\mathbf{0}^T,\mathbf{h}_{m,k}^T]^T,~\text{otherwise},
\end{cases}
\end{eqnarray}
where we adopt the idea of one-hot encoding to embed the information of direct communication or interference links. 
\orange{On the other hand, since $\mathbf{v}_k$ 
is the beamforming variable corresponding to the 
$(m_1(k),k,:)$-th TX-RX pair and $m_1(\cdot)$ is a one-to-one mapping,
we can either define $\mathbf{v}_k$ on the $(m_1(k),k,:)$-th edge, the $k$-th RX-node, or the $m_1(k)$-th TX-node.
Without loss of generality, we put} 
$\mathbf{v}_k$ on the $(m_1(k),k,:)$-th fiber of the edge variable tensor $\boldsymbol{V}$, 
and problem~\eqref{equation-ex1 opt orig} can be reformulated on the heterogeneous graph as 
\neworange{\begin{subequations}\label{equation-ex1 opt}
\begin{align}
&\mathop{\max}\limits_{ \phi\left(\cdot,\cdot,\cdot\right)  }~
\sum^K_{k=1} \log_2 \left(1+\text{SINR}_k\right), \label{equation-ex1 opt loss}\\
    &~\text{s.t.}~~~~ \boldsymbol{V} = \phi\left(\mathbf{p},\bm{\sigma},\boldsymbol{H}\right),~\text{with}~\left\lVert \mathbf{v}_k \right\rVert^2 \leq P_{m_1(k)},~\forall k\in\mathcal{K}. \label{equation-ex1 opt cons2} 
\end{align}
\end{subequations}
Comparing \eqref{equation-ex1 opt} with \eqref{equation-opt}, the objective function
\eqref{equation-ex1 opt loss} is a specification of \eqref{equation-opt_cost}. Particularly, the features $\mathbf{p}$, $\bm{\sigma}$, and $\boldsymbol{H}$ correspond to $\mathbf{F}_{\text{TX}}$, $\mathbf{F}_{\text{RX}}$, and $\boldsymbol{E}$, respectively, and the variable $\boldsymbol{V}$ corresponds to $\boldsymbol{\it{\Xi}}$. }

\emph{\textbf{Example 2: }Power Allocation for Interference Broadcast Channels.}
Consider a $B$-cell downlink cellular network, 
where each BS serves $Q$ UEs. 
Each BS is equipped with $N$ antennas and each UE is equipped with a single antenna. 
The normalized beamforming vector of UE $q\in \mathcal{Q}\triangleq\left\{1,\cdots,Q\right\}$ 
in cell $b\in \mathcal{B}\triangleq\left\{1,\cdots,B\right\}$ 
is denoted as $\mathbf{w}_{q_b}\in \mathbb{C}^N$, 
while the channel between BS $b'$ and UE $q$ in cell $b$ 
is denoted as $\mathbf{h}_{b',q_{b}}\in \mathbb{C}^N$. 
Then the received signal at UE $q$ in cell $b$ is given by 
\begin{equation}
\begin{aligned}
y_{q_b} = \sqrt{p_{q_b}} \mathbf{h}_{b,q_b}^H \mathbf{w}_{q_b} s_{q_b} 
    &+ \sum^Q_{q'=1,q'\neq q}  \sqrt{p_{q'_b}} \mathbf{h}_{b,q_b}^H  \mathbf{w}_{q'_b} s_{q'_b} 
    \\
    &+ \sum^B_{b'=1,b'\neq b} \sum^Q_{q'=1}  \sqrt{p_{q'_{b'}}} \mathbf{h}_{b',q_b}^H  \mathbf{w}_{q'_{b'}} s_{q'_{b'}} 
    + n_{q_b},~\forall q\in\mathcal{Q}, \forall b\in\mathcal{B}, 
\label{equation-exp2 received signal at RX q}
\end{aligned}
\end{equation}
where $s_{q_b}$ and $p_{q_b}$ are the desired symbol and transmit power, 
respectively, 
the second term is the intra-cell interference, the third term is the inter-cell interference, 
and $n_{q_b}\sim\mathcal{C}\mathcal{N}(0,\sigma_{q_b}^2)$ is the additive complex Gaussian noise. 
Correspondingly, the SINR 
is given by 
\begin{equation}
    \text{SINR}_{q_b} =  
    \frac{\left| \mathbf{h}_{b,q_b}^H \mathbf{w}_{q_b} \right|^2 p_{q_b}}
    {\sum^Q_{q'=1,q'\neq q}\left| \mathbf{h}_{b,q_b}^H  \mathbf{w}_{q'_b} \right|^2 p_{q'_b}
    + \sum^B_{b'=1,b'\neq b} \sum^Q_{q'=1}\left| \mathbf{h}_{b',q_b}^H  \mathbf{w}_{q'_{b'}} \right|^2 p_{q'_{b'}}
    + \sigma_{q_b}^2},~\forall q\in\mathcal{Q}, \forall b\in\mathcal{B}. 
\label{equation-ex2 SINR}
\end{equation}
The power allocation problem for sum rate maximization can be formulated as: 
\begin{subequations}\label{equation-ex2 opt orig}
\begin{align}
&\mathop{\max}\limits_{ \left\{ p_{q_b} \right\}_{q\in\mathcal{Q}, b\in\mathcal{B}} }~\sum^B_{b=1}\sum^Q_{q=1}  \log_2 \left(1+ \text{SINR}_{q_b}  \right), \label{equation-ex2 opt loss orig} \\
    &~~~~~\text{s.t.}~~~~~~0 \leq \sum^Q_{q=1} p_{q_b} \leq P_{b},~\forall b\in\mathcal{B}, \label{equation-ex2 opt cons orig}
\end{align}
\end{subequations}
where $P_b$ is the maximum transmit power budget of BS $b$.

As observed in \eqref{equation-ex2 SINR}, the inner product of $\mathbf{h}_{b,q_b}$ and $\mathbf{w}_{q_{b}}$ has $BQ\times BQ$ combinations, leading to $K=BQ$ equivalent TX-nodes and $K$ RX-nodes, respectively. 
By modeling each BS as $Q$ TX-nodes, and each UE as an RX-node, we define two one-to-one mappings $\psi_{\text{TX}}(\cdot,\cdot)$ and $\psi_{\text{RX}}(\cdot,\cdot)$ from $\mathcal{B}\times \mathcal{Q}$ to $\mathcal{M}$ and $\mathcal{K}$, respectively. 
Let $k=\psi_{\text{RX}}(b,q)$, and then for RX$_k$, the TX-nodes can be divided into three types as shown in Fig.~\ref{fig-exs}\subref{fig-ex2}. 
The first type is the TX-node that serves RX$_k$, denoted as TX$_{m_1(k)}$, where $m_1(k)\triangleq\psi_{\text{TX}}(b,q)$. 
The second type consists of other TX-nodes in the same cell as RX$_k$, denoted as TX$_{m_2(k)}$, where $m_2(k) \in \mathcal{M}_2(k)\triangleq\left\{\psi_{\text{TX}}(b,q')|\forall q'\in\mathcal{Q}, q'\neq q \right\}$. 
The third type consists of TX-nodes in other cells, denoted as TX$_{m_3(k)}$, where $m_3(k) \in \mathcal{M}_3(k)\triangleq\left\{\psi_{\text{TX}}(b',q')|\forall b'\in\mathcal{B}, b'\neq b, \forall q'\in\mathcal{Q} \right\}$. 
Accordingly, the equivalent channel gain between different TX-nodes and RX$_k$ can be written as three types: 
\begin{subequations}\label{equation-equivalent channel gain}
\begin{align}
g_{m_1(k),k} &\triangleq \left| \mathbf{h}_{b,q_b}^H  \mathbf{w}_{q_b} \right|, \label{equation-ecg1} \\
g_{m_2(k),k} &\triangleq \left| \mathbf{h}_{b,q_b}^H  \mathbf{w}_{q'_b} \right|,~\forall q'\in\mathcal{Q},q'\neq q, \label{equation-ecg2} \\
g_{m_3(k),k} &\triangleq \left| \mathbf{h}_{b',q_b}^H  \mathbf{w}_{q'_{b'}} \right|,~\forall b'\in\mathcal{B},b'\neq b, \forall q'\in\mathcal{Q}. \label{equation-ecg3}
\end{align}
\end{subequations}

We incorporate the maximum transmit power budget $\mathbf{p} = \left[ \tilde{P}_{1},\cdots,\tilde{P}_{K} \right]^T$ and the noise standard deviation $\bm{\sigma} = [\sigma_1, \cdots, \sigma_K]^T$ as the features on the TX-nodes and RX-nodes, respectively, where 
$\tilde{P}_{m_1(k)} = P_b$. 
Moreover, the feature vector on the $(m,k)$-th edge contains both the equivalent channel gain $g_{m,k}$ and the indicator of direct communication, intra-cell interference, or inter-cell interference link: 
\begin{eqnarray}\label{equation-ex2 graph feature}
\boldsymbol{G}_{(m,k,:)} &=& 
\begin{cases}
    [g_{m,k},0,0]^T,~&\text{if}~m=m_1(k), \cr
    [0,g_{m,k},0]^T,~&\text{if}~m=m_2(k), \cr
    [0,0,g_{m,k}]^T,~&\text{otherwise}, \cr
\end{cases}
\end{eqnarray}
where we adopt the idea of one-hot encoding to embed the information of direct communication, inter-cell interference, or intra-cell interference links. 
\orange{On the other hand, since $p_{q_b}$ 
is the power allocation variable corresponding to the 
$(m_1(k),k,:)$-th TX-RX pair and $m_1(\cdot)$ is a one-to-one mapping,
we can either define $p_{q_b}$ on the $(m_1(k),k,:)$-th edge, the $k$-th RX-node, or the $m_1(k)$-th TX-node.
Without loss of generality, putting} 
the unknown power allocation variable $p_{q_b}$ 
on the $(m_1(k),k,:)$-th fiber of the edge variable tensor $\boldsymbol{P}$, 
problem~\eqref{equation-ex2 opt orig} can be reformulated on the heterogeneous graph as
\neworange{\begin{subequations}\label{equation-ex2 opt}
\begin{align}
&\mathop{\max}\limits_{ \phi\left(\cdot,\cdot,\cdot\right)  }~\sum_{b=1}^B\sum_{q=1}^Q \log_2 \left(1+ \text{SINR}_{q_b}  \right), \label{equation-ex2 opt loss}\\
    &~\text{s.t.}~~~ \boldsymbol{P} = \phi\left(\mathbf{p},\bm{\sigma},\boldsymbol{G}\right),~\text{with}~0 \leq \sum^Q_{q=1} p_{q_b} \leq P_{b},~\forall b\in\mathcal{B}. \label{equation-ex2 opt cons}
\end{align}
\end{subequations}
Comparing \eqref{equation-ex2 opt} with \eqref{equation-opt}, the objective function \eqref{equation-ex2 opt loss} is a specification of \eqref{equation-opt_cost}. In particular, the features $\mathbf{p}$, $\bm{\sigma}$, and $\boldsymbol{G}$ correspond to $\mathbf{F}_{\text{TX}}$, $\mathbf{F}_{\text{RX}}$, and $\boldsymbol{E}$, respectively, and the variable $\boldsymbol{P}$ corresponds to $\boldsymbol{\it{\Xi}}$. }

\emph{\textbf{Example 3: } Cooperative Beamforming Design.} 
Consider a downlink system where $M$ BSs serve $K$ UEs cooperatively. Each BS is equipped with $N$ antennas and serves all UEs, while each UE is equipped with a single antenna and served by all BSs. 
The channel between the $m$-th BS and the $k$-th UE can be defined as $\mathbf{h}_{m,k} \in \mathbb{C}^{N}$. 
The beamforming vector used by the $m$-th BS for serving the $k$-th UE is denoted as $\mathbf{v}_{m,k} \in \mathbb{C}^{N}$. With $s_k$ denoting the desired symbol of the $k$-th UE, the received signal at the $k$-th UE is expressed as
\begin{equation}
    y_k = \sum^M_{m=1}\mathbf{h}^H_{m,k}\mathbf{v}_{m,k} s_{k} + \sum^K_{k'=1,k'\neq k}\sum^M_{m=1}\mathbf{h}^H_{m,k}\mathbf{v}_{m,k'} s_{k'} + n_k,~\forall k\in\mathcal{K},
\label{equation-exp3 received signal at RX k}
\end{equation}
where $n_k\sim\mathcal{CN}(0,\sigma_k^2)$ is the additive complex Gaussian noise. 
The SINR can be written as 
\begin{equation}
    \text{SINR}_k = \frac{\left|\sum^M_{m=1}\mathbf{h}^H_{m,k}\mathbf{v}_{m,k}\right|^2}{\sum^K_{k'=1,k'\neq k}\left|\sum^M_{m=1}\mathbf{h}^H_{m,k}\mathbf{v}_{m,k'}\right|^2+\sigma^2_k},~\forall k\in\mathcal{K}. 
\label{equation-exp3 SINR}
\end{equation}
The cooperative beamforming design problem for sum rate maximization can be formulated as 
\begin{subequations}\label{equation-exp3 sum rate orig}
\begin{align}
&\mathop{\max}\limits_{ \left\{ \mathbf{v}_{m,k} \right\}_{m\in\mathcal{M},k\in\mathcal{K}} }~\sum^K_{k=1} \log_2 \left(1+\text{SINR}_k\right), \label{equation-exp3 sum rate loss orig} \\
    &~~~~~~~\text{s.t.}~~~~~~~\sum^K_{k=1} \left\lVert \mathbf{v}_{m,k} \right\rVert^2 \leq P_{m},~\forall m\in\mathcal{M}, \label{equation-exp3 sum rate cons1 orig}
\end{align}
\end{subequations}
where $P_{m}$ denotes the maximum power budget of BS $m$. 

As illustrated in Fig.~\ref{fig-exs}\subref{fig-ex3}, 
by modeling the BSs and UEs as TX-nodes and RX-nodes respectively, we can incorporate the maximum transmit power budget $\mathbf{p} = [P_1, \cdots, P_M]^T$ and the noise standard deviation $\bm{\sigma} = [\sigma_1, \cdots, \sigma_K]^T$ as the features on the TX-nodes and RX-nodes, respectively. 
Moreover, the feature vector on the $(m,k)$-th edge can be defined as $\boldsymbol{H}_{(m,k,:)} = \mathbf{h}_{m,k}$. 
\orange{Unlike the previous examples, the unknown beamforming variable $\mathbf{v}_{m,k}$ 
is not a variable corresponding to a TX/RX-node, but rather 
corresponding to the $(m,k)$-th TX-RX pair.
Thus, $\mathbf{v}_{m,k}$ can only be defined on the $(m,k)$-th edge.} Putting $\mathbf{v}_{m,k}$ on the $(m,k,:)$-th fiber of the edge variable tensor $\boldsymbol{V}$, 
problem~\eqref{equation-exp3 sum rate orig} can be reformulated on the heterogeneous graph as 
\neworange{\begin{subequations}\label{equation-exp3 sum rate}
\begin{align}
&\mathop{\max}\limits_{ \phi\left(\cdot,\cdot,\cdot\right)  }~\sum^K_{k=1} \log_2 \left(1+\text{SINR}_k\right), \label{equation-exp3 sum rate loss} \\
    &~\text{s.t.}~~~\boldsymbol{V} = \phi\left(\mathbf{p},\bm{\sigma},\boldsymbol{H}\right),~\text{with}~\sum^K_{k=1} \left\lVert \mathbf{v}_{m,k} \right\rVert^2 \leq P_{m},~\forall m\in\mathcal{M}. \label{equation-exp3 sum rate cons1} 
\end{align}
\end{subequations}
Comparing \eqref{equation-exp3 sum rate} with \eqref{equation-opt}, the objective function
\eqref{equation-exp3 sum rate loss} is a specification of \eqref{equation-opt_cost}. Particularly, the features $\mathbf{p}$, $\bm{\sigma}$, and $\boldsymbol{H}$ correspond to $\mathbf{F}_{\text{TX}}$, $\mathbf{F}_{\text{RX}}$, and $\boldsymbol{E}$, respectively, and the variable $\boldsymbol{V}$ corresponds to $\boldsymbol{\it{\Xi}}$.}

Notice that in the cooperative beamforming application, beamforming variables exist on the communication links between TX-nodes and RX-nodes and should therefore be defined on the edges. 
However, existing GNNs \cite{shen2020graph} \cite{guo2021learning} are only equipped with the node-update mechanism, 
which cannot cope with the more complicated problem of cooperative beamforming design, where each TX-node serves multiple RX-nodes and each RX-node is also served by multiple TX-nodes.

\subsection{PE Property}

A unique property of beamforming design and power allocation is 
that the optimized strategy is independent of the indices of TX-nodes and RX-nodes. 
In particular, the learned mapping function $\phi(\cdot,\cdot,\cdot)$ is inherently permutation equivariant with respect to TX-nodes and RX-nodes, i.e., if the indices of any two TX-nodes or RX-nodes are exchanged, $\phi(\cdot,\cdot,\cdot)$ should output a corresponding permutation.

\begin{figure*}[!t]
\centering
\subfloat[Original heterogeneous graph.]{\includegraphics[width=3in]{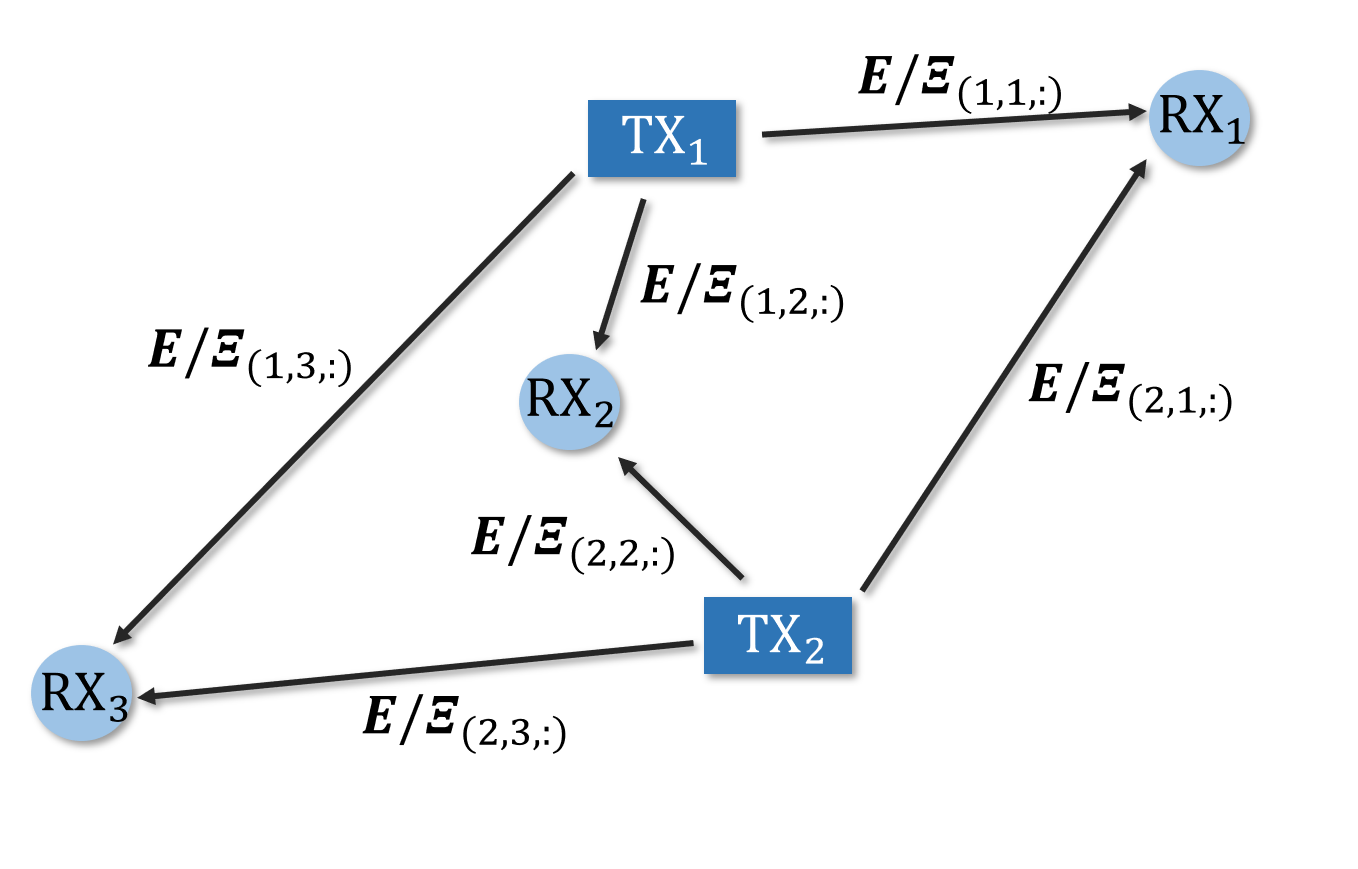}
\label{fig-case1}}
\hfil
\subfloat[Heterogeneous graph after permutations of TX-nodes and RX-nodes.]{\includegraphics[width=3in]{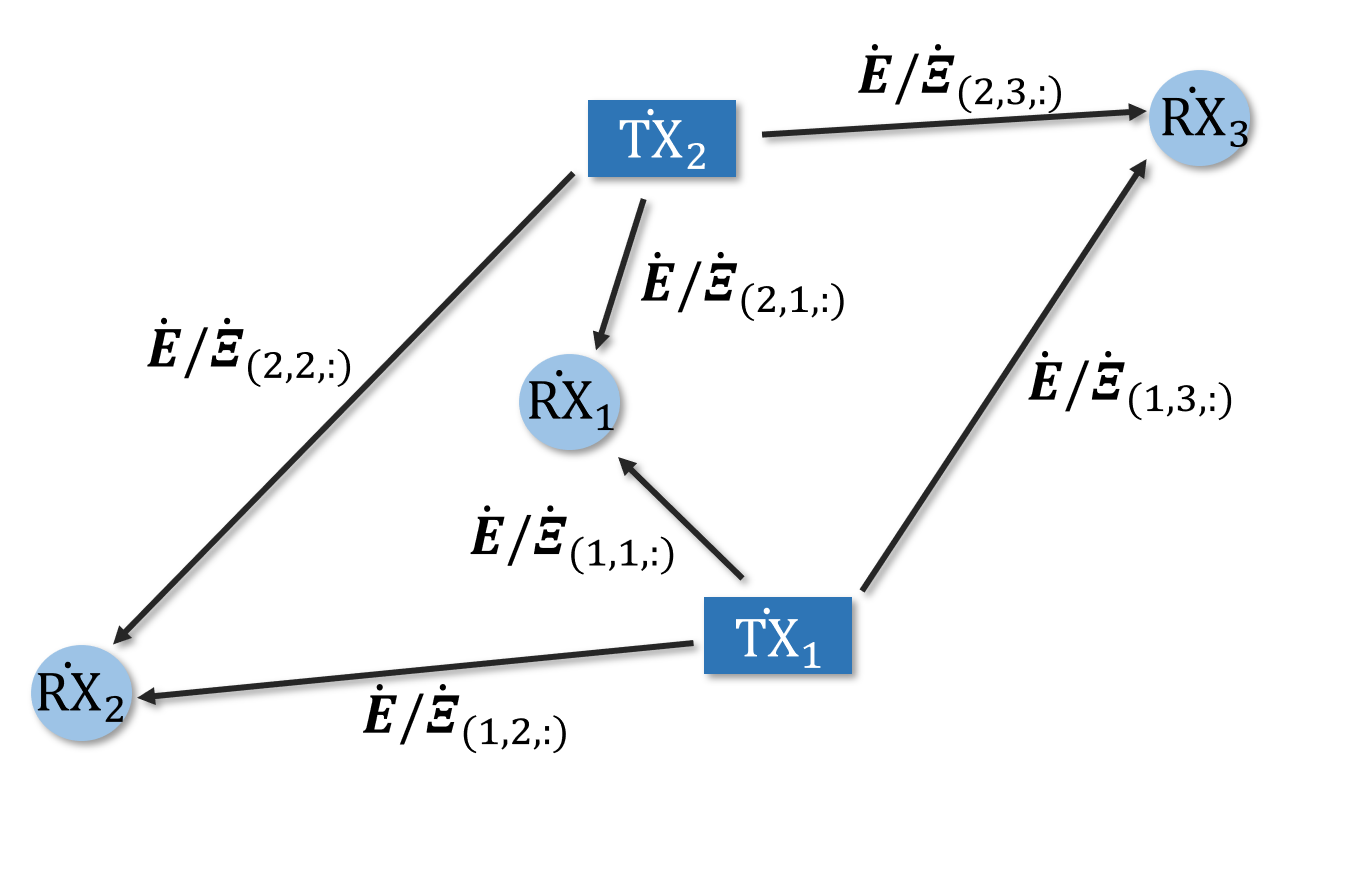}
\label{fig-case2}}
\caption{Permutation equivariance illustration.}
\label{fig-permutation equivariance}
\end{figure*}

To visualize this, 
we show a heterogeneous graph 
before and after permutations of TX-nodes and RX-nodes 
in Fig.~\ref{fig-permutation equivariance}. 
Define two permutations $\pi_{\text{TX}}(\cdot)$ and $\pi_{\text{RX}}(\cdot)$, and let 
${\text{TX}}_m$ and ${\text{RX}}_k$ 
in Fig.~\ref{fig-permutation equivariance}\subref{fig-case1} be re-ordered as 
$\dot{\text{TX}}_{\pi_{\text{TX}}(m)}$ and $\dot{\text{RX}}_{\pi_{\text{RX}}(k)}$ in Fig.~\ref{fig-permutation equivariance}\subref{fig-case2}, 
where $\pi_{\text{TX}}(1)=2$, $\pi_{\text{TX}}(2)=1$, $\pi_{\text{RX}}(1)=3$, $\pi_{\text{RX}}(2)=1$, and $\pi_{\text{RX}}(3)=2$. 
Accordingly, the graph features satisfy 
\begin{subequations}\label{equation-property PE}
\begin{align}
\mathbf{\dot{f}}_{\text{TX},\pi_{\text{TX}}(m)} &= \mathbf{{f}}_{\text{TX},m},~\forall m\in\mathcal{M}, \label{equation-property PE in TX feature}\\
\mathbf{\dot{f}}_{\text{RX},\pi_{\text{RX}}(k)} &= \mathbf{{f}}_{\text{RX},k},~\forall k\in\mathcal{K}, \label{equation-property PE in RX feature}\\
\boldsymbol{\it{\dot{E}}}_{(\pi_{\text{TX}}(m),\pi_{\text{RX}}(k),:)} &= \boldsymbol{\it{{E}}}_{(m,k,:)},~\forall (m,k)\in\mathcal{E}. \label{equation-property PE in edge feature}
\end{align}
\end{subequations}
\orange{Let $\left(\mathbf{\dot{S}}_{\text{TX}}, \mathbf{\dot{S}}_{\text{RX}}, \boldsymbol{\it{\dot{\Xi}}} \right)=\phi\left(\mathbf{\dot{F}}_{\text{TX}},\mathbf{\dot{F}}_{\text{RX}},\boldsymbol{\it{\dot{E}}}\right)$ 
and $\left(\mathbf{{S}}_{\text{TX}}, \mathbf{{S}}_{\text{RX}}, \boldsymbol{\it{{\Xi}}} \right)=\phi\left(\mathbf{{F}}_{\text{TX}},\mathbf{{F}}_{\text{RX}},
\boldsymbol{\it{{E}}}\right)$ 
be the corresponding outputs of the mapping function $\phi(\cdot,\cdot,\cdot)$, respectively.} 
Since $\left(\mathbf{\dot{F}}_{\text{TX}},\mathbf{\dot{F}}_{\text{RX}},\boldsymbol{\it{\dot{E}}}\right)$ is just a re-ordering of the TX-nodes and RX-nodes in $\left(\mathbf{{F}}_{\text{TX}},\mathbf{{F}}_{\text{RX}},
\boldsymbol{\it{{E}}}\right)$, the corresponding outputs of the mapping function $\phi(\cdot,\cdot,\cdot)$ should satisfy
\begin{subequations}\label{equation-pe goal}
\begin{align}
\mathbf{\dot{s}}_{\text{TX},\pi_{\text{TX}}(m)} &= \mathbf{{s}}_{\text{TX},m},~\forall m\in\mathcal{M},  \\
\mathbf{\dot{s}}_{\text{RX},\pi_{\text{RX}}(k)} &= \mathbf{{s}}_{\text{RX},k},~\forall k\in\mathcal{K},  \\
\boldsymbol{\it{\dot{\Xi}}}_{(\pi_{\text{TX}}(m),\pi_{\text{RX}}(k),:)} &= \boldsymbol{\it{{\Xi}}}_{(m,k,:)},~\forall (m,k)\in\mathcal{E}.
\end{align}
\end{subequations}
We will show in the next section that \eqref{equation-pe goal} can be guaranteed by 
the proposed ENGNN with 
properly designed edge/node-update mechanisms. 

% ######################################################################################### %
%                               Section IV     Proposed ENGNN                            %
% ######################################################################################### %
\section{The Proposed ENGNN}                  % ====================================== %
\label{sec:ENGNN}

In this section, we propose 
a customized neural network architecture 
to represent the mapping function $\phi(\cdot,\cdot,\cdot)$
from 
$\left(\mathbf{F}_{\text{TX}}, \mathbf{F}_{\text{RX}}, \boldsymbol{E}\right)$ to 
\orange{$\left(\mathbf{S}_{\text{TX}}, \mathbf{S}_{\text{RX}}, \boldsymbol{\it{\Xi}} \right)$} in problem~\eqref{equation-opt}. 
The proposed neural network architecture incorporates both
an edge-update mechanism and a node-update mechanism
into a GNN and hence we name it as ENGNN. 
We will show that the proposed ENGNN enjoys the PE property given by \eqref{equation-pe goal}. 

\subsection{Overall Architecture}
\begin{figure}[t!]
\centering 
\includegraphics[width=0.8\linewidth]{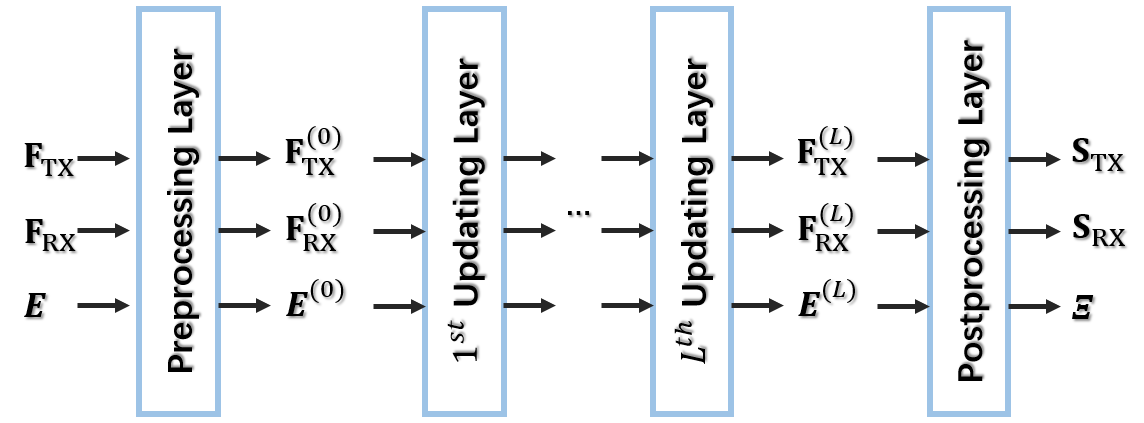}
\caption{\orange{The overall architecture of the proposed ENGNN, which contains a preprocessing layer, $L$ updating layers, and a postprocessing layer.}}
\label{fig-network architecture}
\end{figure}

The proposed ENGNN consists of a preprocessing layer, $L$ updating layers, and a postprocessing layer as illustrated in Fig.~\ref{fig-network architecture}. 
The preprocessing layer transforms the input features $\left(\mathbf{F}_{\text{TX}}, \mathbf{F}_{\text{RX}}, \boldsymbol{E}\right)$ into the initial node- and edge-representations $\left(\mathbf{F}_{\text{TX}}^{(0)}\in\mathbb{R}^{M\times \breve{d}_{\text{TX}}}, \mathbf{F}_{\text{RX}}^{(0)}\in\mathbb{R}^{K\times \breve{d}_{\text{RX}}},  \boldsymbol{E}^{(0)}\in\mathbb{R}^{M\times K\times \breve{d}_{\text{E}}}\right)$, where $\breve{d}_{\text{TX}}$, $\breve{d}_{\text{RX}}$, and $\breve{d}_{\text{E}}$ denote the dimensions of the representations on TX-nodes, RX-nodes, and edges, respectively.
These representations will be updated according to node- and edge-update mechanisms in the $L$ updating layers, where 
the $l$-th updating layer takes $\left(\mathbf{F}_{\text{TX}}^{(l-1)}, \mathbf{F}_{\text{RX}}^{(l-1)},  \boldsymbol{E}^{(l-1)}\right)$ as the inputs and then outputs the updated representations $\left(\mathbf{F}_{\text{TX}}^{(l)}, \mathbf{F}_{\text{RX}}^{(l)},  \boldsymbol{E}^{(l)}\right)$. The dimensions of the representations will not change in the updating layers. 
Finally, the postprocessing layer 
transforms the graph representations $\left(\mathbf{F}_{\text{TX}}^{(L)}, \mathbf{F}_{\text{RX}}^{(L)},  \boldsymbol{E}^{(L)}\right)$ into 
\orange{variables $\left(\mathbf{S}_{\text{TX}}, \mathbf{S}_{\text{RX}}, \boldsymbol{\it{\Xi}} \right)$.}

\subsection{Preprocessing Layer}
The preprocessing layer converts complex-valued features (if any) on TX-nodes, RX-nodes, and edges
into the real-valued form that can be processed by neural networks, and then transforms the real-valued features into initial representations. 
Specifically, the inputs of the preprocessing layer 
$\left(\mathbf{F}_{\text{TX}}, \mathbf{F}_{\text{RX}},  \boldsymbol{E}\right)$
are converted into their corresponding real-valued forms
$\left(\hat{\mathbf{F}}_{\text{TX}}, 
\hat{\mathbf{F}}_{\text{RX}},  \hat{\boldsymbol{E}}\right)$, 
where the $m$-th row of $\mathbf{F}_{\text{TX}}$, the $k$-th row of $\mathbf{F}_{\text{RX}}$, and the $(m,k)$-th fiber of $\boldsymbol{E}$ are given by 
\begin{subequations}\label{equation-preprocessing im2re}
    \begin{align}
    \hat{\mathbf{f}}_{\text{TX},m} &= \left[ \Re\left\{ \mathbf{f}_{\text{TX},m}\right\}^T,  \Im\left\{ \mathbf{f}_{\text{TX},m}\right\}^T \right]^T ,~\forall m \in \mathcal{M}, 
    \label{equation-preprocessing TX im2re} \\
    \hat{\mathbf{f}}_{\text{RX},k} &= \left[ \Re\left\{ \mathbf{f}_{\text{RX},k}\right\}^T,  \Im\left\{ \mathbf{f}_{\text{RX},k}\right\}^T \right]^T ,~\forall k \in \mathcal{K}, 
    \label{equation-preprocessing RX im2re} \\
    \boldsymbol{\hat{E}}_{(m,k,:)} &= \left[ \Re\left\{ \boldsymbol{E}_{(m,k,:)}\right\}^T, \Im\left\{ \boldsymbol{E}_{(m,k,:)}\right\}^T \right]^T ,~\forall (m,k) \in \mathcal{E}. \label{equation-preprocessing E im2re}
    \end{align}
\end{subequations}
Then, the initial representations of TX-nodes, RX-nodes, and edges are transformed by a one-layer MLP with rectified linear unit (ReLU) as the activation function: 
\begin{subequations}\label{equation-preprocessing}
    \begin{align}
    \mathbf{f}_{\text{TX},m}^{(0)} &= \text{ReLU}\left(\mathbf{W}_{\text{TX}}^{\text{pre}}\hat{\mathbf{f}}_{\text{TX},m} + \mathbf{b}_{\text{TX}}^{\text{pre}}\right) ,~\forall m \in \mathcal{M}, \label{equation-preprocessing TX} \\
    \mathbf{f}_{\text{RX},k}^{(0)} &= \text{ReLU}\left(\mathbf{W}_{\text{RX}}^{\text{pre}}\hat{\mathbf{f}}_{\text{RX},k} + \mathbf{b}_{\text{RX}}^{\text{pre}}\right) ,~\forall k \in \mathcal{K}, \label{equation-preprocessing RX} \\
    \boldsymbol{E}^{(0)}_{(m,k,:)} &= \text{ReLU}\left(\mathbf{\mathbf{W}^{\text{pre}}\boldsymbol{\hat{E}}}_{(m,k,:)} + \mathbf{b}^{\text{pre}}\right),~\forall (m,k) \in \mathcal{E}, 
\label{equation-preprocessing E}
\end{align}
\end{subequations}
where 
$\mathbf{W}_{\text{TX}}^{\text{pre}}\in \mathbb{R}^{\breve{d}_{\text{TX}}\times 2d_{\text{TX}}}$, $\mathbf{b}_{\text{TX}}^{\text{pre}}\in \mathbb{R}^{\breve{d}_{\text{TX}} }$, 
$\mathbf{W}_{\text{RX}}^{\text{pre}}\in \mathbb{R}^{\breve{d}_{\text{RX}}\times 2d_{\text{RX}}}$, $\mathbf{b}_{\text{RX}}^{\text{pre}}\in \mathbb{R}^{\breve{d}_{\text{RX}} }$, 
$\mathbf{W}^{\text{pre}}\in \mathbb{R}^{\breve{d}_{\text{ E}}\times 2d_{\text{E}}}$, and $\mathbf{b}^{\text{pre}}\in \mathbb{R}^{\breve{d}_{\text{E}} }$ 
are trainable parameters.

\subsection{Updating Layer}
The inputs and outputs of the updating layer $l\in\left\{1,\cdots,L\right\}$ are 
$\left(\mathbf{F}_{\text{TX}}^{(l-1)}, \mathbf{F}_{\text{RX}}^{(l-1)}, \boldsymbol{E}^{(l-1)}\right)$ 
and 
$\left(\mathbf{F}_{\text{TX}}^{(l)}, \mathbf{F}_{\text{RX}}^{(l)}, \boldsymbol{E}^{(l)}\right)$, respectively. 
We next show the node- and edge-update mechanisms in the $l$-th updating layer.

\subsubsection{Node-Update Mechanism}
\label{sec: node-update mechanism}
The update of node representations in the $l$-th updating layer takes $\left(\mathbf{F}_{\text{TX}}^{(l-1)}, \mathbf{F}_{\text{RX}}^{(l-1)}, \boldsymbol{E}^{(l-1)}\right)$ as the inputs, and then outputs the updated node representations $\left(\mathbf{F}_{\text{TX}}^{(l)}, \mathbf{F}_{\text{RX}}^{(l)}\right)$. 
In particular, when updating the representation of TX$_m$, 
the inputs are composed of the previous layer's representations of
TX$_m$, the neighboring RX-nodes RX$_{k}$, and edges $(m,k)$ for all $k\in\mathcal{N}^{\text{TX}}_m$, where $\mathcal{N}^{\text{TX}}_m$ is 
the set of neighboring RX-nodes of TX$_m$.
First, the input representations on RX$_{k}$ and edge $(m,k)$ 
are concatenated and then processed by an MLP. 
Next, the processing results from all RX$_k$ with $k\in\mathcal{N}^{\text{TX}}_m$
are combined by an aggregation function (e.g., mean or max aggregators), which extracts information from all the neighboring RX-nodes regardless of their input order. 
Finally, the input representation of TX$_m$ and the aggregated result are concatenated and then 
processed by another MLP. 
The above procedure gives the following TX-update mechanism in the $l$-th updating layer:
\begin{equation} \label{equation-EGNN encoding TX}
    \mathbf{f}_{\text{TX},m}^{(l)} = \text{MLP}_2^{(l)}\left( \mathbf{f}_{\text{TX},m}^{(l-1)} , \text{AGG}_{\text{TX}}^{(l)}  \left\{ \text{MLP}_1^{(l)}\left(\mathbf{f}_{\text{RX},k}^{(l-1)}, \boldsymbol{E}_{(m,k,:)}^{(l-1)}\right)  \right\}_{k \in \mathcal{N}^{\text{TX}}_m }\right),~\forall m\in\mathcal{M}, 
\end{equation}
where $\mathbf{f}_{\text{TX},m}^{(l-1)}$ is the $m$-th row of $\mathbf{F}_{\text{TX}}^{(l-1)}$,
$\mathbf{f}_{\text{RX},k}^{(l-1)}$ is the $k$-th row of $\mathbf{F}_{\text{RX}}^{(l-1)}$, 
$\boldsymbol{E}_{(m,k,:)}^{(l-1)}$ is the $(m,k)$-th fiber of $\boldsymbol{E}^{(l-1)}$,
$\text{MLP}_1^{(l)}$ and $\text{MLP}_2^{(l)}$ are two MLPs, and $\text{AGG}_{\text{TX}}^{(l)}$ is an aggregation function.
Similarly, the RX-update mechanism in the $l$-th updating layer 
reverses the roles of TX-nodes and UE-nodes in \eqref{equation-EGNN encoding TX}: 
\begin{equation}\label{equation-EGNN encoding RX}
    \mathbf{f}_{\text{RX},k}^{(l)} = \text{MLP}_4^{(l)}\left( \mathbf{f}_{\text{RX},k}^{(l-1)} , \text{AGG}_{\text{RX}}^{(l)}  \left\{ \text{MLP}_3^{(l)}\left(\mathbf{f}_{\text{TX},m}^{(l-1)}, \boldsymbol{E}_{(m,k,:)}^{(l-1)}\right)  \right\}_{m \in \mathcal{N}^{\text{RX}}_k }\right),~\forall k\in\mathcal{K},
\end{equation}
where $\mathcal{N}^{\text{RX}}_k$ is the set of neighboring TX-nodes of RX$_k$,
$\text{MLP}_3^{(l)}$ and $\text{MLP}_4^{(l)}$ are two MLPs, and $\text{AGG}_{\text{RX}}^{(l)}$ is an aggregation function.

Notice that the representations of TX-nodes and RX-nodes are updated differently in the proposed ENGNN. 
This is different from the previous work MPGNN \cite{shen2020graph} in which 
the node 
representations are updated 
homogeneously.
Moreover, in the proposed ENGNN, the
input edge representations $\boldsymbol{E}_{(m,k,:)}^{(l-1)}$ in \eqref{equation-EGNN encoding TX}
and \eqref{equation-EGNN encoding RX} are 
with superscript $(l-1)$ and hence 
are also updated (see the edge-update mechanism).
Taking a similar analysis 
in \cite{shen2020graph}, we can show that 
\eqref{equation-EGNN encoding TX} and \eqref{equation-EGNN encoding RX} 
satisfy the following PE property: 

$\textbf{Property~1 (PE in Node-Update Mechanism):}$ \emph{The node-update mechanism~\eqref{equation-EGNN encoding TX} and \eqref{equation-EGNN encoding RX} are permutation equivariant with respect to TX-nodes and RX-nodes, respectively. 
Specifically, for any permutations $\pi_{\text{TX}}(\cdot)$ and $\pi_{\text{RX}}(\cdot)$, we have }
\begin{subequations} 
    \begin{align} 
    \mathbf{f}_{\text{TX},\pi_{\text{TX}}(m)}^{(l)} &=& \text{MLP}_2^{(l)}\left( \mathbf{f}_{\text{TX},\pi_{\text{TX}}(m)}^{(l-1)} , \text{AGG}_{\text{TX}}^{(l)}  \left\{ \text{MLP}_1^{(l)}\left(\mathbf{f}_{\text{RX},k}^{(l-1)}, \boldsymbol{E}_{\left(\pi_{\text{TX}}(m),k,:\right)}^{(l-1)}\right)  \right\}_{k \in \mathcal{N}^{\text{TX}}_{\pi_{\text{TX}}(m)} }\right),~\forall m\in\mathcal{M}, 
    \label{PE-BS}
    \end{align}
    \begin{align} 
    \mathbf{f}_{\text{RX},\pi_{\text{RX}}(k)}^{(l)} &=& \text{MLP}_4^{(l)}\left( \mathbf{f}_{\text{RX},\pi_{\text{RX}}(k)}^{(l-1)} , \text{AGG}_{\text{RX}}^{(l)}  \left\{ \text{MLP}_3^{(l)}\left(\mathbf{f}_{\text{TX},m}^{(l-1)}, \boldsymbol{E}_{\left(m,\pi_{\text{RX}}(k),:\right)}^{(l-1)}\right)  \right\}_{m \in \mathcal{N}^{\text{RX}}_{\pi_{\text{RX}}(k)} }\right),~\forall k\in\mathcal{K}.\label{PE-UE}
    \end{align}
\end{subequations}

\subsubsection{Edge-Update Mechanism}
\label{sec: edge-update mechanism}
\begin{figure}[t!]
\centering 
\includegraphics[width=0.8\linewidth]{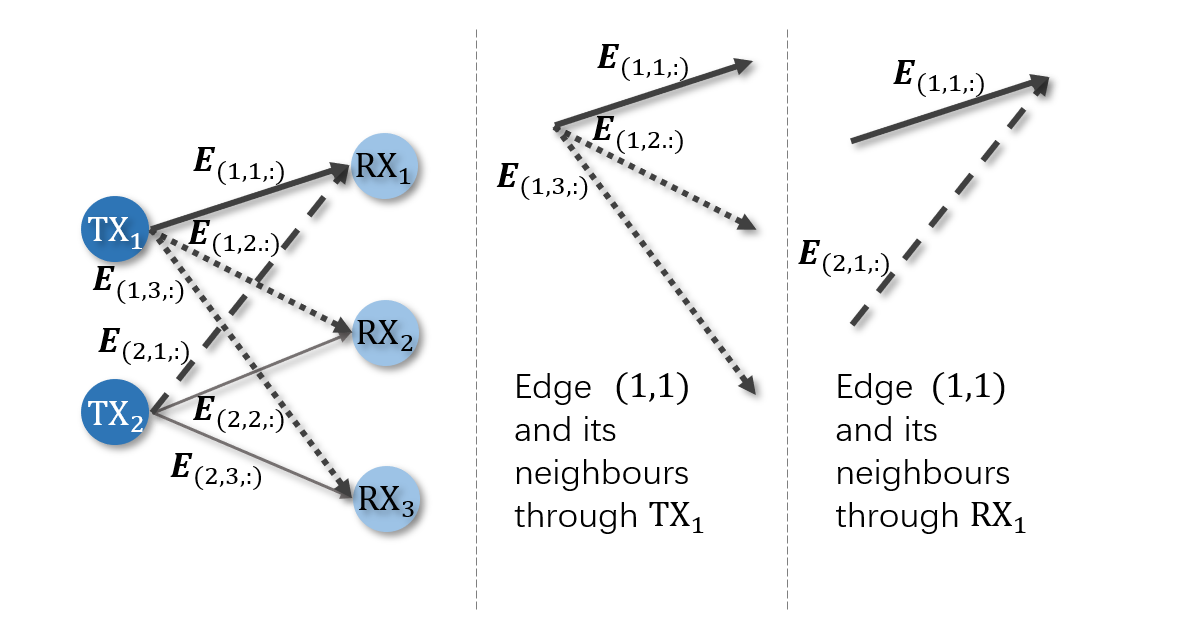}
\caption{The neighbors that share the connection with edge $(1,1)$ through $\text{TX}_1$ are edge $(1,2)$ and edge $(1,3)$, which are denoted by dotted lines. The neighbor that shares the connection with edge $(1,1)$ through $\text{RX}_1$ is edge $(2,1)$, which is denoted by a dashed line.}
\label{fig-edge neighbor}
\end{figure}

The update of edge representations in the $l$-th updating layer takes $\left(\mathbf{F}_{\text{TX}}^{(l-1)}, \mathbf{F}_{\text{RX}}^{(l-1)}, \boldsymbol{E}^{(l-1)}\right)$ as the inputs, and then outputs the updated edge representations $\boldsymbol{E}^{(l)}$.
Different from the 
node-update mechanism, 
where 
the neighbors of a TX-node (or RX-node) are clearly defined as the connecting RX-nodes (or TX-nodes),
it is more complicated to define the neighbors of an edge, let alone how to aggregate 
their representations. 
Notice that an edge may connect with other edges through 
either a TX-node or an RX-node. 
In particular, for the edge $(m,k)\in\mathcal{E}$, 
its neighboring edges through TX$_m$ are $(m,k_1), \forall k_1 \in \mathcal{N}^{\text{TX}}_m \setminus \{k\}$,
while the neighboring edges through RX$_k$
are 
$(m_1,k), \forall m_1 \in \mathcal{N}^{\text{RX}}_k \setminus \{m\}$.
For example, in Fig.~\ref{fig-edge neighbor}, 
the neighboring edges of edge $(1,1)$ through $\text{TX}_1$ are edge $(1,2)$ and edge $(1,3)$. 
On the other hand,
the neighboring edge of edge $(1,1)$ through $\text{RX}_1$ is edge $(2,1)$.
This causes the neighbors of an edge to be innately divided into two categories based on the connecting node.
Consequently, different from the node-update mechanism~\eqref{equation-EGNN encoding TX} and~\eqref{equation-EGNN encoding RX}, 
the edge-update mechanism should provide two different aggregations for the two types of neighboring edges.

Specifically, when updating the representation of edge $(m,k)$, 
the inputs are composed of the previous representations of
edge $(m,k)$, 
TX$_{m}$, 
RX$_{k}$,
the neighboring edges $(m,k_1), \forall k_1 \in \mathcal{N}^{\text{TX}}_m \setminus \{k\}$,
and 
neighboring edges 
$(m_1,k), \forall m_1 \in \mathcal{N}^{\text{RX}}_k \setminus \{m\}$.
The input representations of neighboring edges $(m,k_1), \forall k_1 \in \mathcal{N}^{\text{TX}}_m \setminus \{k\}$ and the connecting node TX$_{m}$ are concatenated and then processed by an MLP,
while the input representations of neighboring edges $(m_1,k), \forall m_1 \in \mathcal{N}^{\text{RX}}_k \setminus \{m\}$ and the connecting node RX$_{k}$ are concatenated and then processed by another MLP. 
The processing results of all the neighboring edges are aggregated and then
concatenated with the input representation of edge $(m,k)$.
An MLP is finally applied to produce the updated representation of edge $(m,k)$.
We can express the above edge-update procedure in the $l$-th updating layer as
\begin{equation}
\begin{aligned}
    \boldsymbol{E}_{(m,k,:)}^{(l)} =~&\text{MLP}_7^{(l)}\biggr( \boldsymbol{E}_{(m,k,:)}^{(l-1)} , \text{AGG}_{\text{E}}^{(l)}  \left\{ \text{MLP}_5^{(l)}\left(\boldsymbol{E}_{(m,k_1,:)}^{(l-1)}, \mathbf{f}_{\text{TX},m}^{(l-1)}\right) , \right.  \\
    &\left. \text{MLP}_6^{(l)}\left(\boldsymbol{E}_{(m_1,k,:)}^{(l-1)}, \mathbf{f}_{\text{RX},k}^{(l-1)}\right)  \right\}_{k_1 \in \mathcal{N}^{\text{TX}}_m \setminus \{k\}, m_1 \in \mathcal{N}^{\text{RX}}_k \setminus \{m\}}\biggr),~\forall (m,k)\in\mathcal{E},
\label{equation-EGNN encoding edge}
\end{aligned}
\end{equation}
where $\boldsymbol{E}_{(m,k,:)}^{(l-1)}$ is the $(m,k)$-th fiber of $\boldsymbol{E}^{(l-1)}$,
$\text{MLP}_5^{(l)}$, $\text{MLP}_6^{(l)}$, and $\text{MLP}_7^{(l)}$ are three MLPs, 
and $\text{AGG}_{\text{E}}^{(l)}$ is an aggregation function.

Compared with the node-update mechanism~\eqref{equation-EGNN encoding TX} and ~\eqref{equation-EGNN encoding RX}, the edge-update mechanism~\eqref{equation-EGNN encoding edge} is more complicated, 
since the definition of neighbors in edge-update mechanism is more complex than that in the node-update one. 
In particular, 
the 
edge-update mechanism 
faces a more complicated situation where the neighboring edges are innately divided into two categories according to the two possible connected nodes. 
Consequently, different from the 
node-update mechanism~\eqref{equation-EGNN encoding TX} and~\eqref{equation-EGNN encoding RX}, where the information from neighboring nodes are gathered by one MLP, 
the proposed edge-update mechanism applies two different 
transformations to extract the information from two different types of neighboring edges.
We next show that \eqref{equation-EGNN encoding edge} enjoys the following PE property: 

$\textbf{Property~2 (PE in Edge-Update Mechanism):}$ \emph{The edge-update mechanism~\eqref{equation-EGNN encoding edge} is permutation equivariant with respect to TX-nodes and RX-nodes. 
Specifically, for any permutations $\pi_{\text{TX}}(\cdot)$ and $\pi_{\text{RX}}(\cdot)$, we have }
\begin{equation} 
\begin{aligned}
    &\boldsymbol{E}_{(\pi_{\text{TX}}(m),\pi_{\text{RX}}(k),:)}^{(l)} =~\text{MLP}_7^{(l)}\left( \boldsymbol{E}_{(\pi_{\text{TX}}(m),\pi_{\text{RX}}(k),:)}^{(l-1)} , \text{AGG}_{\text{E}}^{(l)}  \left\{ \text{MLP}_5^{(l)}\left(\boldsymbol{E}_{(\pi_{\text{TX}}(m),k_1,:)}^{(l-1)}, \mathbf{f}_{\text{TX},\pi_{\text{TX}}(m)}^{(l-1)}\right) , \right.\right.  \\
    &\left.\left. \text{MLP}_6^{(l)}\left(\boldsymbol{E}_{(m_1,\pi_{\text{RX}}(k),:)}^{(l-1)}, \mathbf{f}_{\text{RX},\pi_{\text{RX}}(k)}^{(l-1)}\right)  \right\}_{k_1 \in \mathcal{N}^{\text{TX}}_{\pi_{\text{TX}}(m)} \setminus \left\{\pi_{\text{RX}}(k)\right\}, m_1 \in \mathcal{N}^{\text{RX}}_{\pi_{\text{RX}}(k)} \setminus \left\{\pi_{\text{TX}}(m)\right\}}\right),~\forall (m,k)\in\mathcal{E}.  \label{equation-PE-edge}
\end{aligned}
\end{equation}

\textit{Proof:} See Appendix \ref{appendix-property 2}.

\subsection{Postprocessing Layer}

The postprocessing layer converts the graph representations $\left(\mathbf{F}_{\text{TX}}^{(L)}, \mathbf{F}_{\text{RX}}^{(L)},  \boldsymbol{E}^{(L)}\right)$ into the final output $\left(\mathbf{S}_{\text{TX}}, \mathbf{S}_{\text{RX}}, \boldsymbol{\it{\Xi}} \right)$. 
First, if the variables are complex, 
$\left(\mathbf{F}_{\text{TX}}^{(L)}, \mathbf{F}_{\text{RX}}^{(L)},  \boldsymbol{E}^{(L)}\right)$ are transformed into 
the complex form $\left(\mathbf{\tilde{S}}_{\text{TX}}, \mathbf{\tilde{S}}_{\text{RX}}, \boldsymbol{\it{\tilde{\Xi}}} \right)$ by
\begin{subequations}\label{equation-postprocessing layer bm}
    \begin{align}
    \left[\Re\left\{\mathbf{\tilde{s}}_{\text{TX},m}\right\}^T, \Im\left\{\mathbf{\tilde{s}}_{\text{TX},m}\right\}^T\right]^T &= \mathbf{W}_{\text{TX}}^{\text{post}}\mathbf{f}_{\text{TX},m}^{(L)} + \mathbf{b}_{\text{TX}}^{\text{post}},~\forall m \in \mathcal{M},  \\
    \left[\Re\left\{\mathbf{\tilde{s}}_{\text{RX},k}\right\}^T, \Im\left\{\mathbf{\tilde{s}}_{\text{RX},k}\right\}^T\right]^T  &= \mathbf{W}_{\text{RX}}^{\text{post}}\mathbf{f}_{\text{RX},k}^{(L)} + \mathbf{b}_{\text{RX}}^{\text{post}},~\forall k \in \mathcal{K},  \\
    \left[\Re\left\{\boldsymbol{\it{\tilde{\Xi}}}_{(m,k,:)}\right\}^T,
     \Im\left\{\boldsymbol{\it{\tilde{\Xi}}}_{(m,k,:)}\right\}^T\right]^T &= \mathbf{W}^{\text{post}}\boldsymbol{E}_{(m,k.:)}^{(L)} + \mathbf{b}^{\text{post}},~\forall (m,k) \in \mathcal{E}, 
\end{align}
\end{subequations}
where $\mathbf{W}_{\text{TX}}^{\text{post}}\in \mathbb{R}^{2d'_{\text{TX}}\times \breve{d}_{\text{TX}}}$, 
$\mathbf{b}_{\text{TX}}^{\text{post}}\in \mathbb{R}^{2d'_{\text{TX}} }$, 
$\mathbf{W}_{\text{RX}}^{\text{post}}\in \mathbb{R}^{2d'_{\text{RX}}\times \breve{d}_{\text{RX}}}$, 
$\mathbf{b}_{\text{RX}}^{\text{post}}\in \mathbb{R}^{2d'_{\text{RX}} }$, 
$\mathbf{W}^{\text{post}}\in \mathbb{R}^{2d'_{\text{E}}\times \breve{d}_{\text{E}}}$, and $\mathbf{b}^{\text{post}}\in \mathbb{R}^{2d'_{\text{E}}}$ are trainable parameters. 
Next, $\left(\mathbf{\tilde{S}}_{\text{TX}}, \mathbf{\tilde{S}}_{\text{RX}}, \boldsymbol{\it{\tilde{\Xi}}} \right)$ are normalized to satisfy the constraints (if any), 
obtaining the final output $\left(\mathbf{S}_{\text{TX}}, \mathbf{S}_{\text{RX}}, \boldsymbol{\it{\Xi}} \right)$.

\subsection{Key Insights}

The proposed ENGNN for representing
$\phi(\cdot, \cdot,  \cdot)$ has been specified as
a preprocessing layer, $L$ updating layers, and a postprocessing layer, where the preprocessing and postprocessing layers utilize edge/node-wise MLPs, and the $L$ updating layers are built on node- and edge-update mechanisms \eqref{equation-EGNN encoding TX}, \eqref{equation-EGNN encoding RX}, and \eqref{equation-EGNN encoding edge}. Next, we provide some key insights of the proposed ENGNN for learning the beamforming design and power allocation as follows. 

\subsubsection{Permutation Equivariant with Respect to TX-nodes and RX-nodes}

$  $ \par
$\textbf{Proposition 1 (PE in ENGNN):}$ \emph{The proposed ENGNN is permutation equivariant with respect to TX-nodes and RX-nodes.
Specifically,
for any permutations $\pi_{\text{TX}}(\cdot)$ and $\pi_{\text{RX}}(\cdot)$, denote a permuted problem instance of $\left(\mathbf{{F}}_{\text{TX}},\mathbf{{F}}_{\text{RX}},\boldsymbol{\it{{E}}}\right)$ as
$\left(\mathbf{{\dot{F}}}_{\text{TX}},\mathbf{{\dot{F}}}_{\text{RX}},\boldsymbol{\it{{\dot{E}}}}\right)$, 
whose entries satisfy \eqref{equation-property PE}.
The corresponding outputs of
the proposed ENGNN, 
\orange{$\left(\mathbf{\dot{S}}_{\text{TX}}, \mathbf{\dot{S}}_{\text{RX}}, \boldsymbol{\it{\dot{\Xi}}} \right)=\phi\left(\mathbf{{\dot{F}}}_{\text{TX}},\mathbf{{\dot{F}}}_{\text{RX}},\boldsymbol{\it{{\dot{E}}}}\right)$} and 
\orange{$\left(\mathbf{S}_{\text{TX}}, \mathbf{S}_{\text{RX}}, \boldsymbol{\it{\Xi}} \right)
=\phi\left(\mathbf{{F}}_{\text{TX}},\mathbf{{F}}_{\text{RX}},\boldsymbol{\it{{E}}}\right)$}, 
always satisfy \eqref{equation-pe goal}.}

\textit{Proof:} See Appendix \ref{appendix-theorem}.

Proposition 1 implies that the proposed ENGNN is
inherently incorporated with the PE 
property. This is in sharp contrast to the generic MLPs,
which require all permutations of each training sample
to approximate this property. Thus, the proposed
ENGNN can reduce the sample complexity and
training difficulty.

\subsubsection{Generalization on Different Numbers of TX-nodes and RX-nodes}
\neworange{In all the layers of the proposed ENGNN, 
the representations on different edges/TX-nodes/RX-nodes are 
transformed by the same architecture using the same trainable parameters.
Therefore, the dimensions of the trainable parameters are independent of the numbers of TX-nodes and RX-nodes.}
This scale adaptability empowers ENGNN
to be trained in a setup with a small graph size, 
while being deployed to a much larger wireless network for the inference.

\subsubsection{Tackling Edge Variables}
The proposed ENGNN is equipped with an edge-update mechanism,
which facilitates the update of the variables on graph edges.
This allows ENGNN to be applied
in a wider range of scenarios, where variables are defined between a pair of nodes.

% % % ######################################################################################### %
% % %                                Section V   Simulation Results                             %
% % % ######################################################################################### %
\section{Simulation Results}
\label{sec: simulation-results}
In this section, we demonstrate the superiority of the proposed ENGNN on the three examples introduced in Section~\ref{sec:examples of the problems} via simulations. 
We consider a downlink wireless network 
in a $2 \times 2$ km$^2$ area, where the BSs and UEs are uniformly distributed. 
Each BS has a maximum transmit power of $33$ dBm. 
The path loss is $30.5 + 36.7 \log_{10}d$ in dB, where $d$ is the distance in meters. 
The small scale channels follow Rayleigh fading and the noise power is $-99$ dBm.

For the proposed ENGNN,
each aggregation function in~\eqref{equation-EGNN encoding TX}, \eqref{equation-EGNN encoding RX}, and \eqref{equation-EGNN encoding edge} is implemented by a max aggregator, which returns the element-wise maximum value of the inputs. 
All the MLPs in~\eqref{equation-EGNN encoding TX}, \eqref{equation-EGNN encoding RX}, and \eqref{equation-EGNN encoding edge} are implemented by $3$ linear layers, each followed by a ReLU activation function. 
In the training procedure, the number of epochs is set to $500$. 
Each epoch consists of $100$ mini-batches of training samples with a batch size of $256$. 
For each training sample, the BSs' and UEs' locations, and the small scale channels are randomly generated.
A learning rate $\gamma=10^{-4}$ is adopted to update the trainable parameters of ENGNN by maximizing \eqref{equation-opt_cost} using RMSProp \cite{tieleman2012lecture} in an unsupervised manner.
After training, we test the average performance of $100$ samples. 
All the experiments are implemented using Pytorch on one NVIDIA V100 GPU ($32$ GB, SMX$2$).

\subsection{Beamforming Design for Interference Channels}

\begin{figure*}[t]
\centering
\includegraphics[width=3in]{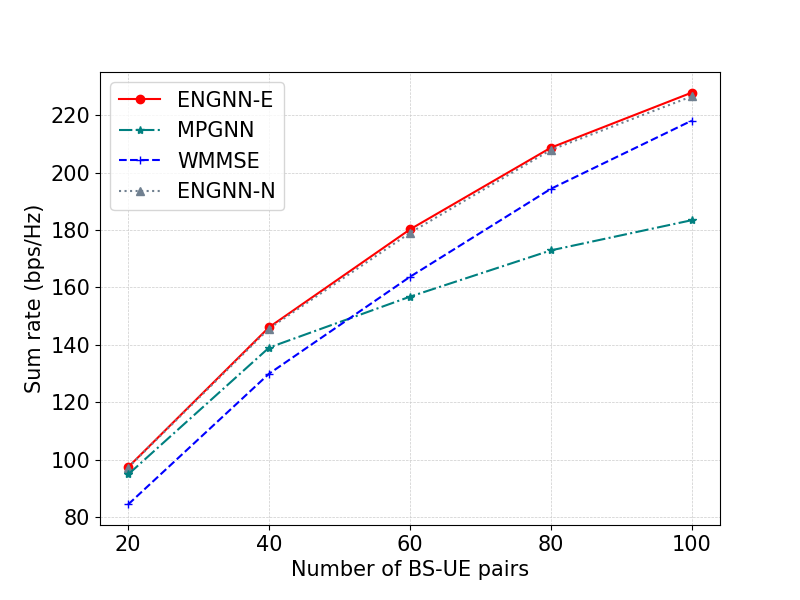}
\caption{Generalization on number of BS-UE pairs.}
\label{ex1-bsue-pairs}
\end{figure*}

First, we demonstrate the performance of the proposed ENGNN on the problem of beamforming design for interference channels. 
During the training procedure, we set the wireless network with $20$ BS-UE pairs, where each BS is equipped with $2$ antennas. 
An ENGNN with $1$ updating layer is adopted, and the dimension of edge features $\breve{d}_{\text{E}}$ is set to $8$. 
As explained before \eqref{equation-ex1 opt}, the beamforming variables in this scenario 
can be defined on either the edges or the nodes,
which only affects the postprocessing layers.
The corresponding simulation results are termed as ENGNN-E and ENGNN-N, respectively. 
For performance comparison, we include two state-of-the-art methods: 
\begin{enumerate}[label=(\alph*)]
\item WMMSE: a widely used benchmark algorithm for sum rate maximization~\cite{shi2011iteratively}.
\item MPGNN: the latest learning based method for sum rate maximization in interference channels~\cite{shen2020graph}. 
\end{enumerate}

\subsubsection{Generalization on Number of BS-UE Pairs}
We first compare the performance of different approaches as the number of BS-UE pairs increases. 
In particular, ENGNN-E, ENGNN-N, MPGNN are trained on $20$ BS-UE pairs, while we test their performance in terms of sum rate on larger problem scales from $20$ to $100$ BS-UE pairs in Fig.~\ref{ex1-bsue-pairs}. 
It can be seen that ENGNN-E, ENGNN-N, and MPGNN generalize well as the number of BS-UE pairs increases from $20$ to $40$. 
However, as the number of BS-UE pairs increases from $40$ to $100$,
both ENGNN-E and ENGNN-N can still generalize very well, 
while the performance of MPGNN becomes worse than that of WMMSE. 
The superiority of ENGNN-E and ENGNN-N is owing to the proposed edge-update mechanism, which better 
extracts the features from channel states and hence further empowers the original node-update mechanism in MPGNN. 
Since ENGNN-E and ENGNN-N result in similar performance, we only show
ENGNN-E in the rest of simulations and term it as ENGNN for simplicity.

\subsubsection{Generalization on Noise Power}

\begin{figure*}[t]
\centering
\includegraphics[width=3in]{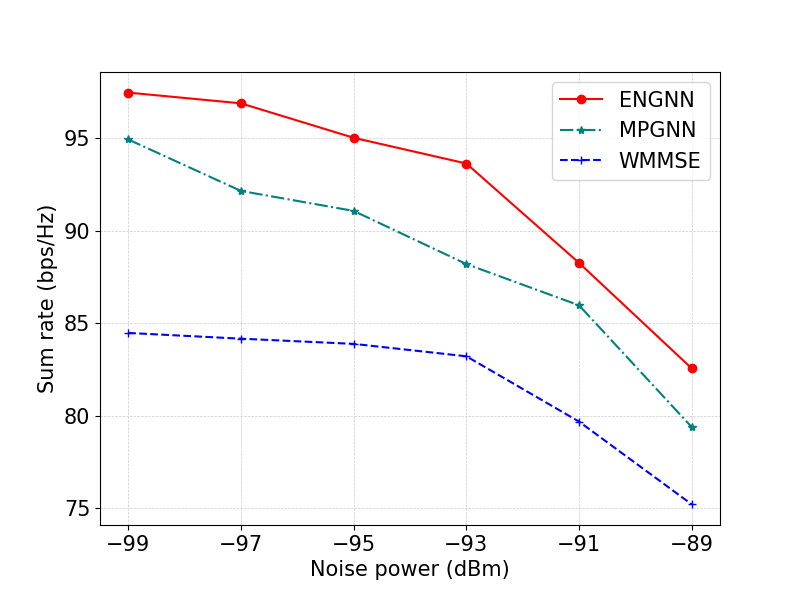}
\caption{Generalization on noise power.}
\label{ex1-noise}
\end{figure*}

To demonstrate the generalization performance on noise power, we set the noise power during the training procedure as $-99$ dBm, while we test the generalization performance under different noise powers from $-99$ dBm to $-89$ dBm. 
It can be seen from Fig.~\ref{ex1-noise} that ENGNN consistently achieves higher sum rate than those of MPGNN and WMMSE, which demonstrates the superiority of ENGNN in generalizing to different noise powers.

\begin{figure*}[t]
\centering
\includegraphics[width=3in]{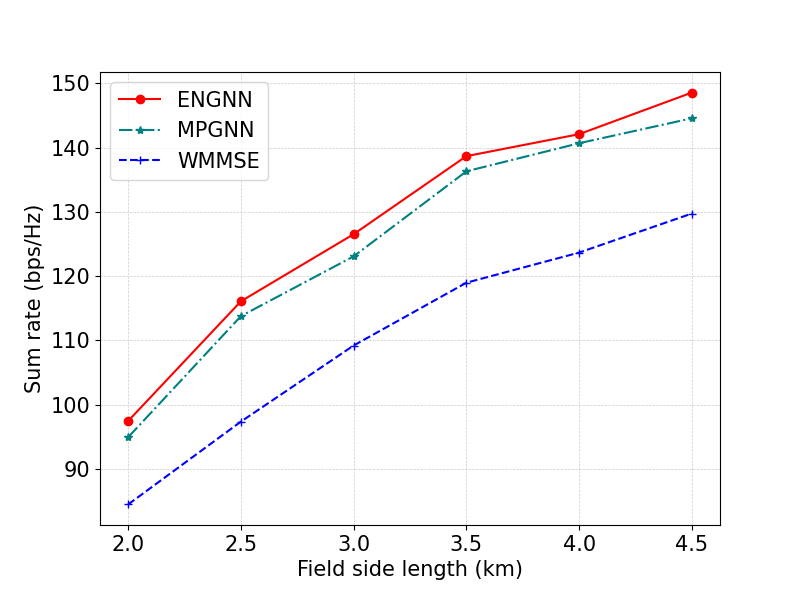}
\caption{Generalization on different levels of interference among BS-UE pairs.}
\label{ex1-d}
\end{figure*}

\begin{figure*}[t]
\centering
\includegraphics[width=3in]{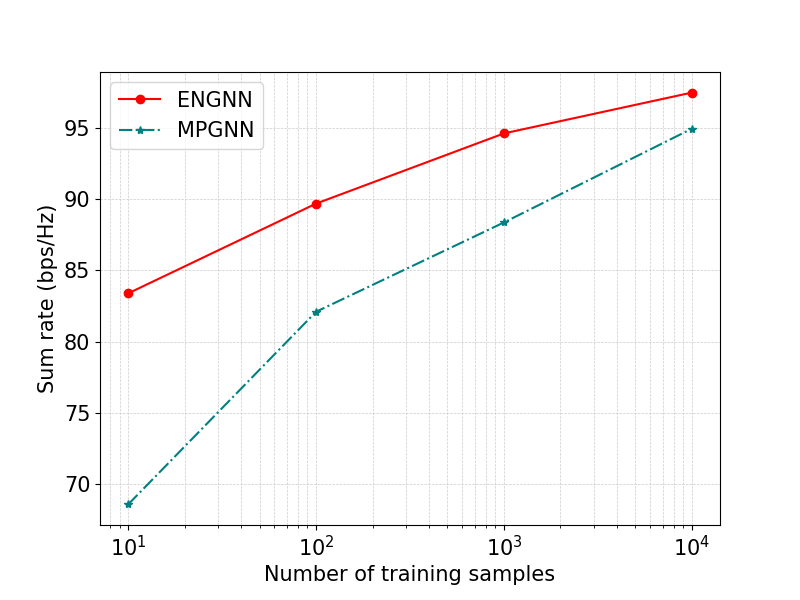}
\caption{Sample complexity comparison between ENGNN and MPGNN.}
\label{ex1-trainingsamples}
\end{figure*}

\subsubsection{Generalization on Different Levels of Interference}

In this experiment, we fix the field size during the training procedure as 
$2 \times 2$ km$^2$, while we test the performance by varying the field size from $2 \times 2$ km$^2$ to $4.5 \times 4.5$ km$^2$. 
By keeping the distance between each BS and its serving UE within $50$-$250$ meters, the interference among different BS-UE pairs becomes weaker as the field size increases, and hence the sum rates of different approaches become higher in Fig.~\ref{ex1-d}.
Moreover, ENGNN consistently achieves higher sum rate than those of MPGNN and WMMSE as the field size increases, which demonstrates that ENGNN generalizes well on different levels of interference.

\subsubsection{Sample Complexity Comparison}

We further compare the performance of ENGNN and MPGNN when trained on different numbers of training samples 
in Fig.~\ref{ex1-trainingsamples}.
It can be seen that ENGNN outperforms MPGNN especially when the number of training samples is small. 
In particular, when the number of training samples decreases to $10$-$100$, the sum rate of ENGNN only decreases to $83.39$-$89.68$ bps/Hz, while that of MPGNN decreases sharply to $68.60$-$82.10$ bps/Hz. 
The required number of training samples of ENGNN is less than $10\%$ of that of MPGNN when they achieve the same sum rate.
This demonstrates the advantage of the proposed edge-update mechanism in sample complexity.

\subsection{Power Allocation for Interference Broadcast Channels}

Next, we demonstrate the performance of ENGNN on the problem of power allocation for interference broadcast channels. 
In the training procedure, we set the wireless network with $5$ BSs, with a minimum distance of $500$ meters between BSs. 
Each BS is equipped with $16$ antennas and serves $2$ UEs. 
We use zero-forcing beamforming to avoid multi-user interference. 
The ENGNN is set with $1$ updating layer and a dimension of $\breve{d}_{\text{E}}=32$ for the edge features. 
For performance comparison, we provide the simulation results of 
WMMSE and PGNN~\cite{guo2021learning}, which is the latest learning-based method for power allocation in interference broadcast channels. 

\subsubsection{Generalization on Number of UEs}

\begin{figure*}[t]
\centering
\includegraphics[width=3in]{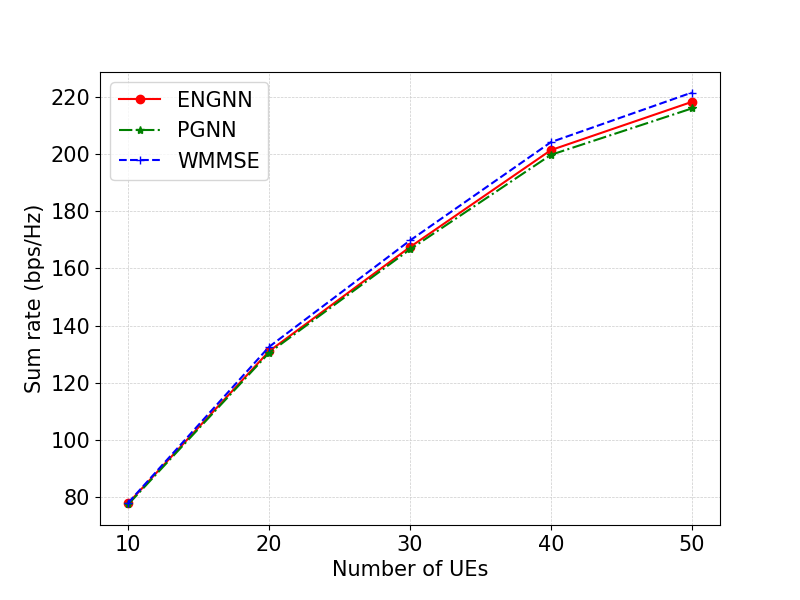}
\caption{Generalization on number of UEs for power allocation in interference broadcast channels.}
\label{ex2-ue}
\end{figure*}

We first compare the performance of different approaches as the number of UEs increases. 
In particular, both ENGNN and PGNN are trained under $10$ UEs, 
while we test their performance 
on larger problem scales from $10$ to $50$ UEs by varying the number of UEs in each cell from $2$ to $10$ in Fig.~\ref{ex2-ue}. 
It can be seen that both ENGNN and PGNN generalize well as the number of UEs increases from $10$ to $50$. 
However, 
ENGNN always outperforms PGNN, and achieves competitive performance compared to that of WMMSE under different numbers of UEs.

\subsubsection{Generalization on Different Levels of Interference}

\begin{figure*}[t]
\centering
\includegraphics[width=3in]{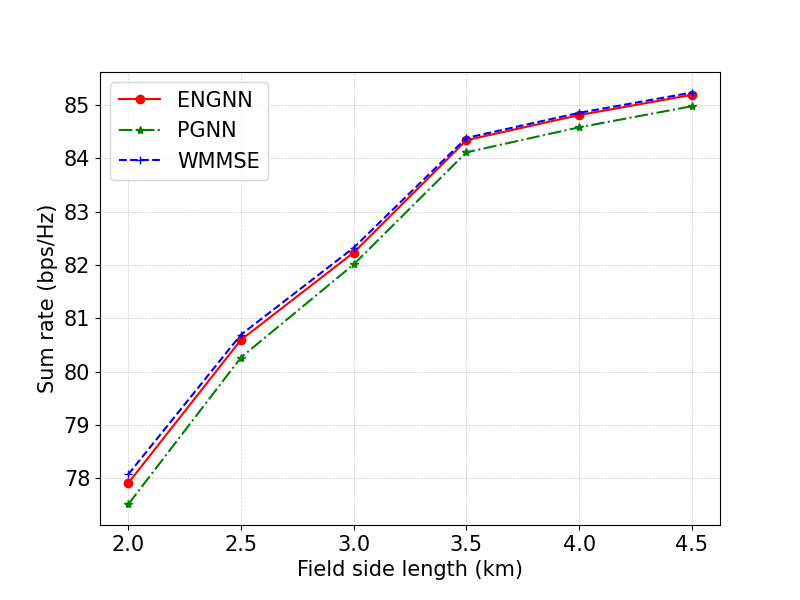}
\caption{Generalization on different levels of inter-cell interference.}
\label{ex2-d}
\end{figure*}

\begin{figure*}[t]
\centering
\includegraphics[width=3in]{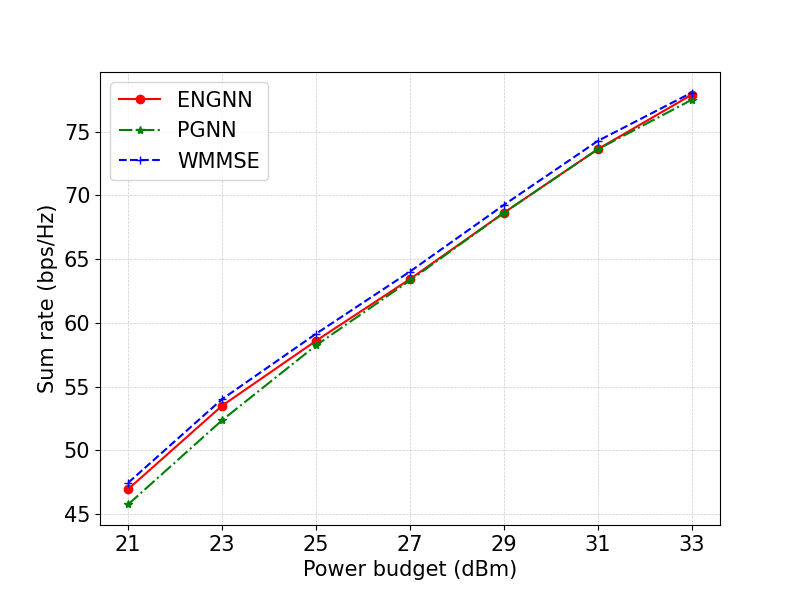}
\caption{Generalization on power budget.}
\label{ex2-pmax}
\end{figure*}

To demonstrate the generalization ability on different levels of interference, during the training procedure, the field size is fixed as $2 \times 2$ km$^2$, while we test the performance by varying the field size from $2 \times 2$ km$^2$ to $4.5 \times 4.5$ km$^2$. 
By keeping the distance between each BS and its serving UE within $50$-$250$ meters, the inter-cell interference becomes weaker as the field size increases, and hence the sum rate becomes higher in Fig.~\ref{ex2-d}. 
It can be observed that the advantage of ENGNN over PGNN is stable as the field size increases, which demonstrates its superior generalization capability on different levels of interference.

\subsubsection{Generalization on Power Budget}

To demonstrate the generalization capability of ENGNN on different power budgets, we set
the power budget at each BS as $33$ dBm during the training procedure, while we test the generalization performance under different power budgets from $21$ dBm to $33$ dBm in Fig.~\ref{ex2-pmax}. 
It can be seen that both ENGNN and PGNN generalize well as the power budgets decreases from $33$ dBm to $21$ dBm. 
However, ENGNN always outperforms PGNN under different power budgets.

\begin{figure*}[t]
\centering
\includegraphics[width=3in]{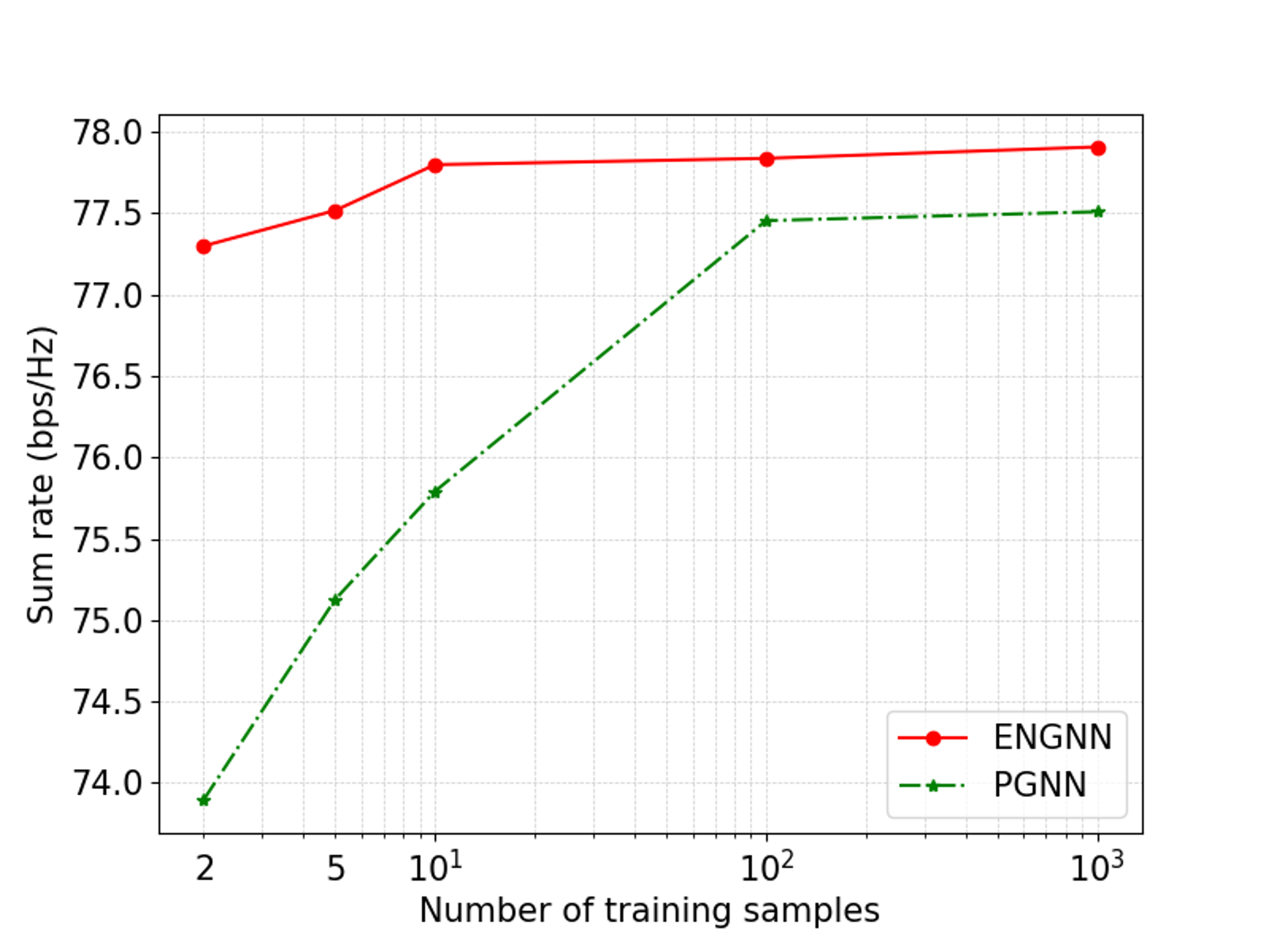}
\caption{Sample complexity comparison between ENGNN and PGNN.}
\label{ex2-trainingsamples}
\end{figure*}

\subsubsection{Sample Complexity Comparison}

We further compare the performance of ENGNN and PGNN when trained on different numbers of samples in Fig.~\ref{ex2-trainingsamples}. 
It can be seen that ENGNN outperforms PGNN especially when the number of training samples is small. 
Particularly, when the number of training samples decreases to $2$-$5$, the sum rate of ENGNN only drops to $77.30$-$77.52$ bps/Hz, while that of PGNN drops to $73.89$-$75.12$ bps/Hz. 
The required number of training samples of ENGNN is about $1\%$ of that of MPGNN when they achieve the same sum rate. 
This demonstrates the advantage of the proposed edge-update mechanism 
in sample complexity.

\subsection{Cooperative Beamforming Design}

Finally, we demonstrate the performance of ENGNN on the problem of cooperative beamforming design. 
During the training procedure, we set the wireless network with $5$ BSs and $2$ UEs. 
Each BS is equipped with $2$ antennas and the minimum distance between BSs is $500$ m. 
An ENGNN with $2$ updating layers is adopted, and the dimension of edge features $\breve{d}_{\text{E}}$ is set to $64$. 
For performance comparison, we include WMMSE and GP~\cite{bertsekas1997nonlinear}, the latter being a computationally efficient first-order algorithm for solving simply constrained optimization problems.

\begin{figure}[t]
\centering
\subfloat[Sum rate comparison]{\includegraphics[width=3in]{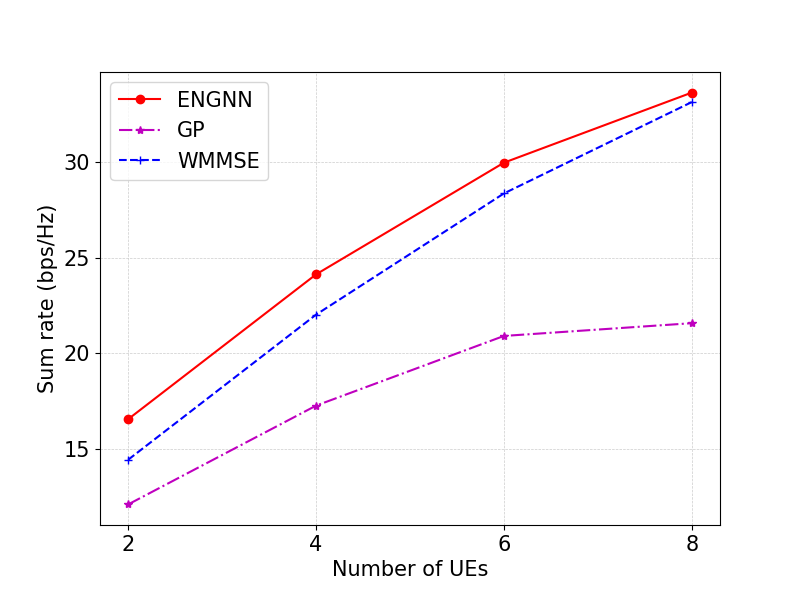}
\label{fig-exp3 RX sr}}
\hfil
\subfloat[Computation time comparison]{\includegraphics[width=3in]{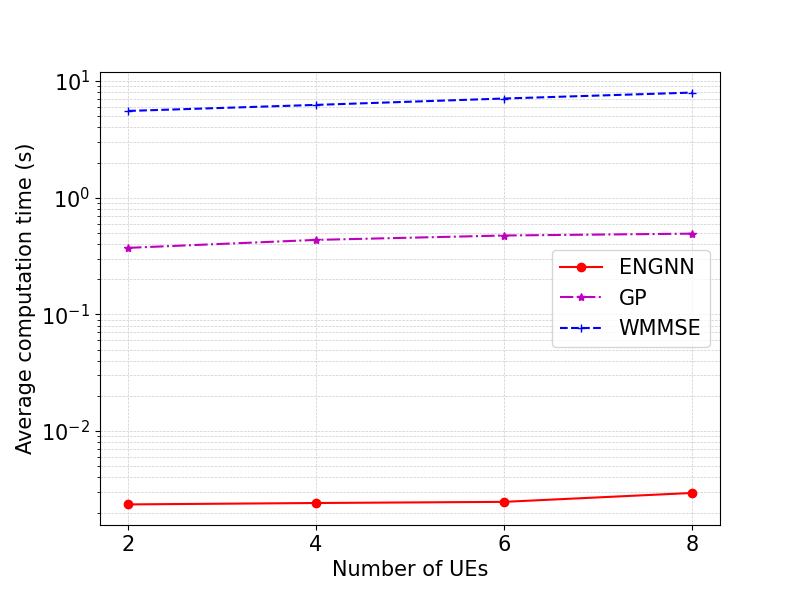}
\label{fig-exp3 RX time}}
\caption{Generalization on number of UEs for cooperative beamforming.}
\label{fig-exp3 RX}
\end{figure}

\subsubsection{Generalization on Number of UEs}
To demonstrate the generalization ability of ENGNN with respect to different numbers of UEs, 
during the training procedure, the number of UEs is fixed as $2$,
while we test the performance of the trained ENGNN by varying the number of UEs from $2$ to $8$. 
The performance comparison in terms of sum rate and computation time is shown in Fig.~\ref{fig-exp3 RX}.
We observe from Fig.~\ref{fig-exp3 RX}\subref{fig-exp3 RX sr}
that as the number of UEs increases, ENGNN always 
outperforms GP and WMMSE in terms of sum rate, which 
demonstrates its generalization ability with respect to different numbers of UEs. 
On the other hand, Fig.~\ref{fig-exp3 RX}\subref{fig-exp3 RX time} shows that ENGNN achieves a remarkable running speed, with over $100$ times faster than that of GP and over $1000$ times faster than that of WMMSE due to the computationally efficient feed forward computations.

\begin{figure}[t]
\centering
\subfloat[Sum rate comparison]{\includegraphics[width=3in]{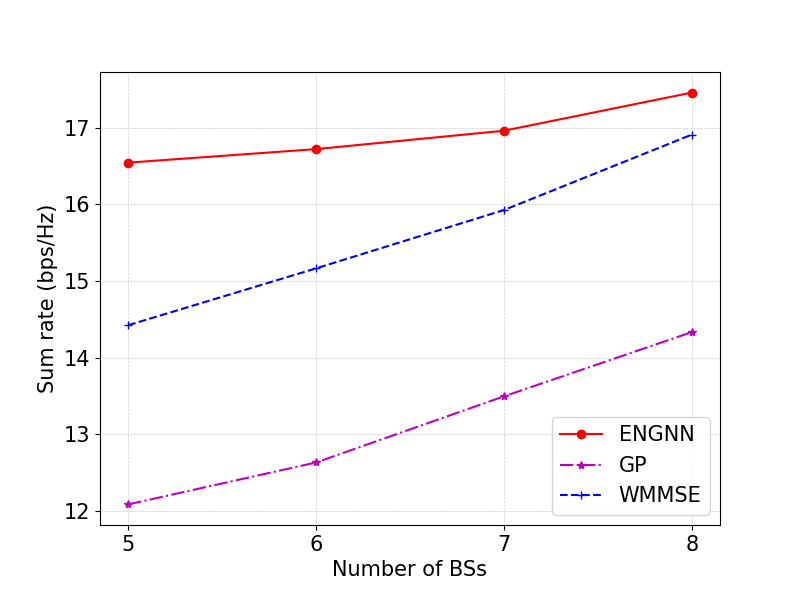}
\label{fig-TX sr}}
\hfil
\subfloat[Computation time comparison]{\includegraphics[width=3in]{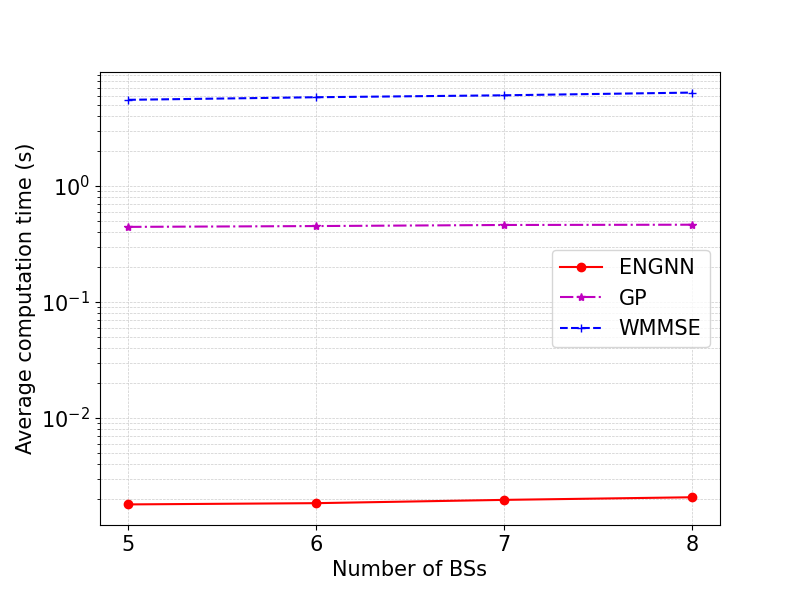}
\label{fig-TX time}}
\caption{Generalization on numbers of BSs for cooperative beamforming.}
\label{fig-exp1 TX}
\end{figure}

\subsubsection{Generalization on Number of BSs}
We further demonstrate the generalization ability of ENGNN with respect to different numbers of BSs. 
Specifically, the number of BSs is fixed as $5$ during the training procedure,
while we test the performance of the trained ENGNN by varying the number of BSs from $5$ to $8$. 
The performance comparison is shown in Fig.~\ref{fig-exp1 TX}.
We observe from Fig.~\ref{fig-exp1 TX}\subref{fig-TX sr} that ENGNN achieves higher sum rate than those of GP and WMMSE under different numbers of BSs. 
Moreover, Fig.~\ref{fig-exp1 TX}\subref{fig-TX time} shows that 
ENGNN achieves a much faster running speed 
than that of GP and 
WMMSE under different numbers of BSs.

\section{Conclusion}
\label{sec: concluion}
This paper proposed a general problem formulation on the heterogeneous graph for the radio resource management problems, where the unknown variables to be designed can be defined on both the graph nodes and edges.
A novel edge-update mechanism with desired PE property was incorporated, and a general neural network architecture 
was designed based on it, which can represent the mapping function from the node/edge features to variables 
for the radio resource management problems.
Simulation results demonstrated the superiority of the proposed architecture on three typical problems, with higher sum rate and much shorter computation time compared with state-of-the-art methods. 
Moreover, the proposed architecture generalizes well on different numbers of BSs and UEs, different noise variances, interference levels, and transmit power budgets.

\appendices

\section{Proof of Property 2}
\label{appendix-property 2}
Let $m'=\pi_{\text{TX}}(m)$ and $k'=\pi_{\text{RX}}(k)$. Substituting these two equations into \eqref{equation-EGNN encoding edge}, we have 
\begin{equation}
\begin{aligned}
    \boldsymbol{E}_{(m',k',:)}^{(l)} =~&\text{MLP}_7^{(l)}\left( \boldsymbol{E}_{(m',k',:)}^{(l-1)} , \text{AGG}_{\text{E}}^{(l)}  \left\{ \text{MLP}_5^{(l)}\left(\boldsymbol{E}_{(m',k'_1,:)}^{(l-1)}, \mathbf{f}_{\text{TX},m'}^{(l-1)}\right) , \right.\right.  \\
    &\left.\left. \text{MLP}_6^{(l)}\left(\boldsymbol{E}_{(m'_1,k',:)}^{(l-1)}, \mathbf{f}_{\text{RX},k'}^{(l-1)}\right)  \right\}_{k'_1 \in \mathcal{N}^{\text{TX}}_{m'} \setminus \{k'\}, m'_1 \in \mathcal{N}^{\text{RX}}_{k'} \setminus \{m'\}}\right),~\forall (m',k')\in\mathcal{E},
\label{equation-edge pe proof}
\end{aligned}
\end{equation}
which implies that for any 
$\pi_{\text{TX}}(\cdot)$ and $\pi_{\text{RX}}(\cdot)$, we always have \eqref{equation-PE-edge}.

\section{Proof of Proposition 1}
\label{appendix-theorem}

Substituting \eqref{equation-property PE} into \eqref{equation-preprocessing im2re} and \eqref{equation-preprocessing}
we have 
\begin{subequations}\label{permute}
\begin{align}
\mathbf{\dot{f}}_{\text{TX},\pi_{\text{TX}}(m)}^{(0)} &= \mathbf{f}_{\text{TX},m}^{(0)},~\forall m\in\mathcal{M}, \label{permute TX pre2} \\
\mathbf{\dot{f}}_{\text{RX},\pi_{\text{RX}}(k)}^{(0)} &= \mathbf{f}_{\text{RX},k}^{(0)},~\forall k\in\mathcal{K},\label{permute RX pre2} \\
\boldsymbol{\dot{E}}^{(0)}_{(\pi_{\text{TX}}(m),\pi_{\text{RX}}(k),:)} &= \boldsymbol{E}^{(0)}_{(m,k,:)},~\forall (m,k)\in\mathcal{E}. 
\label{permute edge pre2}
\end{align}
\end{subequations}
Next, we substitute \eqref{permute} into \eqref{equation-EGNN encoding TX}, \eqref{equation-EGNN encoding RX}, and 
\eqref{equation-EGNN encoding edge}. 
According to \textbf{Property 1} and \textbf{Property 2}, we have
\begin{subequations}\label{permute updating}
\begin{align}
\mathbf{\dot{f}}_{\text{TX},\pi_{\text{TX}}(m)}^{(l)} &= \mathbf{f}_{\text{TX},m}^{(l)},~\forall m\in\mathcal{M}, \forall l=1,\cdots,L,\label{permute TX updating} \\
\mathbf{\dot{f}}_{\text{RX},\pi_{\text{RX}}(k)}^{(l)} &= \mathbf{f}_{\text{RX},k}^{(l)},~\forall k\in\mathcal{K}, \forall l=1,\cdots,L,\label{permute RX updating} \\
\boldsymbol{\dot{E}}^{(l)}_{(\pi_{\text{TX}}(m),\pi_{\text{RX}}(k),:)} &= \boldsymbol{E}_{(m,k.:)}^{(L)},~\forall (m,k)\in\mathcal{E}, \forall l=1,\cdots,L.  
\label{permute edge updating}
\end{align}
\end{subequations}
Substituting \eqref{permute updating} into \eqref{equation-postprocessing layer bm}, we obtain
\begin{subequations}
\begin{align}
    \mathbf{\dot{\tilde{s}}}_{\text{TX},\pi_{\text{TX}}(m)} &= \mathbf{\tilde{s}}_{\text{TX},m},
    ~\forall m \in \mathcal{M},
    \label{equation-hat TX postprocessing} \\
    \mathbf{\dot{\tilde{s}}}_{\text{RX},\pi_{\text{RX}}(k)} &= \mathbf{\tilde{s}}_{\text{RX},k},
    ~\forall k \in \mathcal{K},
    \label{equation-hat RX postprocessing} \\
   \boldsymbol{\it{\dot{\tilde{\Xi}}}}_{(\pi_{\text{TX}}(m),\pi_{\text{RX}}(k),:)} &= \boldsymbol{\it{\tilde{\Xi}}}_{(m,k,:)},
    ~\forall (m,k) \in \mathcal{E}.
    \label{equation-hat V bm postprocessing} 
\end{align}
\end{subequations}
Finally, 
since the normalization of $\left(\mathbf{\tilde{S}}_{\text{TX}}, \mathbf{\tilde{S}}_{\text{RX}}, \boldsymbol{\it{\tilde{\Xi}}} \right)$ is an edge/node-wise computation, the final outputs satisfy 
$\mathbf{\dot{s}}_{\text{TX},\pi_{\text{TX}}(m)} = \mathbf{{s}}_{\text{TX},m},~\forall m \in \mathcal{M}$, 
$\mathbf{\dot{s}}_{\text{RX},\pi_{\text{RX}}(k)} = \mathbf{{s}}_{\text{RX},k},~\forall k \in \mathcal{K}$, and 
$\boldsymbol{\it{\dot{\Xi}}}_{(\pi_{\text{TX}}(m),\pi_{\text{RX}}(k),:)} 
= \boldsymbol{\it{\Xi}}_{(m,k,:)},~\forall (m,k) \in \mathcal{E}$.

% if have a single appendix:
%\appendix[Proof of the Zonklar Equations]
% or
% \appendix  % for no appendix heading
% do not use \section anymore after \appendix, only \section*
% is possibly needed

% use appendices with more than one appendix
% then use \section to start each appendix
% you must declare a \section before using any
% \subsection or using \label (\appendices by itself
% starts a section numbered zero.)
%

% \section
% \section{Proof of the First Zonklar Equation}
% Appendix one text goes here.

% % you can choose not to have a title for an appendix
% % if you want by leaving the argument blank
% \section{}
% Appendix two text goes here.

% % use section* for acknowledgment
% \section*{Acknowledgment}

% The authors would like to thank...

% Can use something like this to put references on a page
% by themselves when using endfloat and the captionsoff option.
\ifCLASSOPTIONcaptionsoff
  \newpage
\fi

% yunqi -------------------------
\bibliographystyle{IEEEtran}
\bibliography{IEEEabrv,egbib}

\end{sloppypar}
% that's all folks
\end{document}